\documentclass[11pt,a4paper]{article}
\usepackage{a4wide}
\usepackage{amsmath}
\usepackage{amssymb}
\usepackage{graphicx}
\usepackage{young}
\usepackage{ifxetex}
\ifxetex 
\usepackage[SlantFont]{xeCJK}
\setCJKfamilyfont{tt}{Hiragino Mincho Pro}
\usepackage{hyperref}

\else
\usepackage{CJK}
\usepackage[CJKbookmarks]{hyperref}
 
\makeatletter
\AtBeginDvi{\pdfmapline{=ipamp@Unicode@  <ipamp-select.ttf}}
\DeclareFontFamily{C70}{ipamp}{\hyphenchar \font\m@ne}
\DeclareFontShape{C70}{ipamp}{l}{n}{ <-> CJK * ipamp}{}
\DeclareFontShape{C70}{ipamp}{m}{n}{ <-> CJK * ipamp}{\CJKnormal}
\DeclareFontShape{C70}{ipamp}{bx}{n}{ <-> CJKb * ipamp}{\CJKbold}
\makeatother

\fi

\numberwithin{equation}{section}

\def\bC{\mathbb{C}}
\def\bR{\mathbb{R}}
\def\bZ{\mathbb{Z}}
\def\cA{\mathcal{A}}
\def\cH{\mathcal{H}}
\def\cM{\mathcal{M}}
\def\cN{\mathcal{N}}
\def\SU{\mathrm{SU}}
\def\U{\mathrm{U}}
\def\SO{\mathrm{SO}}
\def\vev#1{\langle#1\rangle}
\def\ket#1{|#1\rangle}

\def\tr{\mathop{\mathrm{tr}}\nolimits}
\def\diag{\mathrm{diag}}
\def\vol{\mathrm{vol}}

\begin{document}
\ifxetex
\else
\begin{CJK}{UTF8}{ipamp}
\fi

\def\figurename{図}
\def\contentsname{目次}

\ifxetex\else 
\begin{titlepage}
\begin{flushright}
IPMU11-0147
\end{flushright}
\vbox{}\vfill
\begin{center}
\def\thefootnote{\fnsymbol{footnote}}
{\Large \bfseries
A strange relationship between 2d CFT and 4d gauge theory
}

\vskip 1.2cm
Yuji Tachikawa
\vskip 1.2cm

IPMU, University of Tokyo,\\
5-1-5 Kashiwa-no-ha, Kashiwa, Chiba, 277-8583 Japan

\vskip 1.5cm

\textbf{abstract}\footnote[1]{The review is prepared in Japanese as is customary for the proceedings of this summer school series, which has more than 20 years of history. An interested reader can find how to post \TeX\ files written in CJK(Chinese-Japanese-Korean) languages to the arXiv by downloading the source code. 
}
\end{center}

A  relationship between 4d gauge theory and 2d CFT will be reviewed from the very basics. 
We will first cover the introductory material on the 2d CFT and on the instantons of 4d gauge theory.
Next we will explicitly calculate and check the agreement of the norm of a coherent state on the 2d side and the instanton partition function on the 4d side.
We will then see how this agreement can be understood from the perspective of string and M theory.

\bigskip\bigskip

\begin{center}
\itshape to appear on the proceedings of the \\[.3em]
``Summer School on Mathematical Physics 2011'', Komaba
\end{center}

\bigskip

\vfill\vfill\vbox{}
\end{titlepage}
\fi

\begin{titlepage}
\vbox{}
\begin{flushright}
IPMU11-0147
\end{flushright}
\vfill
\begin{center}
\def\thefootnote{\fnsymbol{footnote}}
{\Large \bfseries
四次元ゲージ理論と二次元共形場理論の不思議な関係 
}

\vskip 1.2cm
立川\  裕二
\vskip 1.2cm

〒277-8583 千葉県柏市柏の葉 5-1-5 \\
東京大学数物連携宇宙研究機構(IPMU)

\vskip 1.5cm

\textbf{概要}\footnote[1]{これは東大駒場で8月25日から28日にかけて開かれた「数理物理2011」で行った講演のための予稿の誤植を修正したものです。}
\end{center}

近年発見された、四次元ゲージ理論と二次元共形場理論の関係を非常に基礎から解説します。まず、二次元の共形場理論および四次元のインスタントンの入門をした後、
二次元側でコヒーレント状態のノルムの計算、四次元側でインスタントンの分配関数を具体的に計算を行って確認します。その後、一致が超弦理論/M理論の枠内でどのように自然に解釈されるかを見ます。

\vfill\vfill\vbox{}
\end{titlepage}

\tableofcontents
\addtocounter{section}{-1}
\section{はじめに}
超弦理論は、現時点では厳密に数学的に構築されてもいませんし、実験的に物理的に確認されたわけでもありません。しかし、いろいろな状況証拠から判断するに、超弦理論が自己矛盾の無い数学的対象として存在するのは間違いないと思われます\footnote{我々が生まれてくるのが二百年遅ければ、普通に教科書に定義が載っていて不思議ではないだろう、ということです。解析学だって Newton, Leibnitz のころはいろいろ問題がありました。}。

超弦理論は、数学的には非常に大雑把には、10次元の多様体 $M$ に対して分配関数と呼ばれる複素数 $Z(M)$ を対応させる手続きです。この手続きは完全には未だ定義されていませんので、弦理論屋は、これをまず何とか数学的にきちんと定義され、計算できる量まで変形します。しばしば、$Z(M)$ を計算できる量にする方法は何通りもあります。それが二つあるとし、$Z_A$、 $Z_B$ と呼びましょう。単に同じものを二通りの仕方で計算しただけですから、
$Z_A$ と $Z_B$ は等しいはずです(図\ref{sketch})。$Z_A$ も $Z_B$ も数学的にきちんと定義できる量ですから、$Z_A=Z_B$ は現在の数学で扱える主張です。しかし、超弦理論 $Z(M)$ は未だ満足いくように定義されていませんから、しばしば $Z_A=Z_B$ は数学者には何故成立するのか俄には判らない主張として現れることになります。

\begin{figure}\[
\includegraphics[width=.5\textwidth]{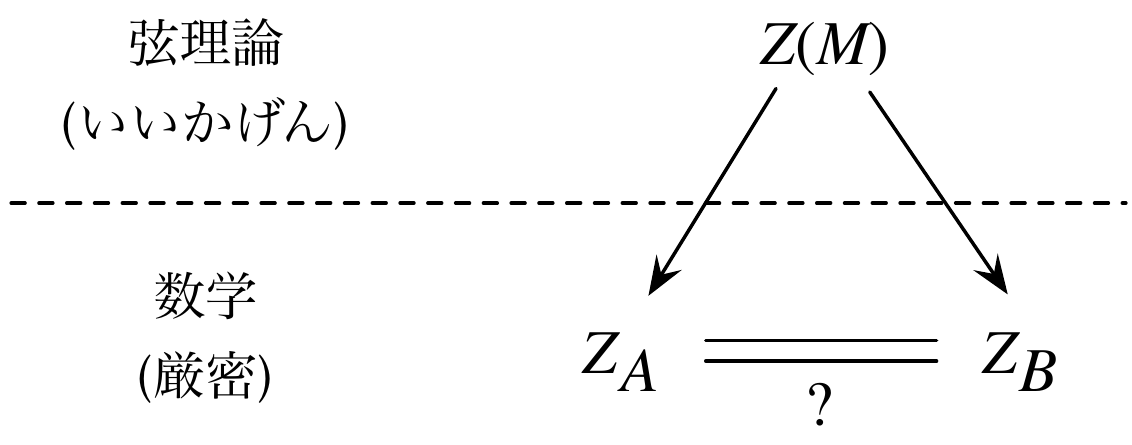}
\]
\caption{弦理論からの数学的現象の導出\label{sketch}}
\end{figure}

このような現象の例の最も著名で、非常に深いものとして90年代初頭に発見されたミラー対称性があげられます。また、Donaldson 不変量と Seiberg-Witten 不変量の一致が物理で云う Seiberg-Witten 理論から従うのもこの現象の一例と思うことができます。さらに、ゲージ・重力対応からも箙の構造と佐々木アインシュタイン多様体の幾何が対応するなど数学的現象を取りだすことが出来ます。

今回の講義では、以上挙げた例に比べるとかなり浅いですが、手を動かして計算しやすい
\begin{align}
Z_A &= \text{二次元共形場理論のコヒーレント状態のノルム}\hbox{、} \\
Z_B &= \text{四次元インスタントンの分配関数}
\end{align} 
の二つが一致するという事実を説明します。$Z(M)$ は、六次元の $\cN=(2,0)$ 理論と呼ばれるものになります。具体的には、$Z_A$ は $c$ を中心電荷、$\Delta$ を $L_0$ の固有値、$\lambda$ をコヒーレント状態のパラメタとして
\begin{equation}
Z_A=\vev{\Delta,\lambda|\Delta,\lambda}=1+\frac{\lambda^2}{2\Delta}+\frac{\lambda^4(c+8\Delta)}{4\Delta((1+\Delta)c-10\Delta+16\Delta^2)}+\cdots \hbox{、}
\end{equation}
$Z_B$ は $q$, $\epsilon_{1,2}$, $a$ をそれぞれインスタントン数、角運動量、ゲージ荷に対する化学ポテンシャルとして、\begin{multline}
Z_B=Z^\text{instanton}_{\epsilon_1,\epsilon_2,a}=1+ \frac{q}{\epsilon_1\epsilon_2} \frac{2}{(\epsilon_1+\epsilon_2)^2 -4a^2}  \\
+ \frac{q^2}{\epsilon_1^2\epsilon_2^2} \frac{( 8(\epsilon_1+\epsilon_2)^2 +\epsilon_1\epsilon_2 - 8a^2)}
{ ((\epsilon_1+\epsilon_2)^2-4a^2)
((2\epsilon_1+\epsilon_2)^2-4a^2) ((\epsilon_1+2\epsilon_2)^2-4a^2) } + \cdots
\end{multline}となります。これらが \begin{equation}
\lambda^2=\frac{q}{(\epsilon_1\epsilon_2)^2}\hbox{、} \qquad
\Delta=\frac1{\epsilon_1\epsilon_2}(\frac{(\epsilon_1+\epsilon_2)^2}{4} -a^2) ,\qquad
c=1+6\frac{(\epsilon_1+\epsilon_2)^2}{\epsilon_1\epsilon_2}
\end{equation} と対応付けると完全に一致する、というものです。

講義の順序は、まずは超弦理論のことはすっかり忘れて、$Z_A$ および $Z_B$ をかなり基礎的なところから説明し、上でつかった用語をまず説明します。次に、$Z_A$ と $Z_B$ を具体的に計算して一致する事を確認したのち、それがどのように自然に超弦理論から理解できるかを説明する、というようにしたいと思います。
このノートは昨年書きました \cite{Gakkaishi} の増補版になっておりますので、もっと短いものをお好みの方はそちらをご参照ください。

\section{二次元共形場理論とそのコヒーレント状態}
\subsection{二次元理論の例}
まずは二次元の数理物理からはじめましょう。(この節の詳細は、例えば教科書 \cite{ID,KY,Yamada,ItoText} を参照のこと。)

二次元の物理系の簡単な例として、イジング模型を考えます。この模型は、二次元面の各格子点 $(i,j)$ に自由度 $\sigma_{i,j}=\pm 1$ があり、ある配位の実現される確率は $T$ を正の数として \begin{equation}
\frac1Z \exp \left[\frac1{T} \sum_{i,j} (\sigma_{i,j}\sigma_{i+1,j} + \sigma_{i,j}\sigma_{i,j+1})\right]
\end{equation} である、というものです。$Z$ は全体の確率が1になるための規格化の定数です。明らかに、\begin{equation}
Z=\sum_{\sigma_{i,j}} \exp \left[\frac1{T} \sum_{i,j} (\sigma_{i,j}\sigma_{i+1,j} + \sigma_{i,j}\sigma_{i,j+1})\right] \label{isingZ}
\end{equation} です。$Z$ を分配関数と呼びます。
$\sigma$ をスピンと呼び、$\sigma=+1$ を上向き、$\sigma=-1$ を下向きといいましょう。すると、隣り合った格子点 $p$ と $q$ に対して、スピン $\sigma_p$ と $\sigma_q$ が揃っているほうが揃っていない場合より $\exp(2/T)$ だけ可能性が高いわけです。
$T$ を温度と呼びましょう。この用語は次のように考える事ができます。$T=0$ では、隣り合った格子点のスピンは揃う確率が揃わない確率に対して無限に可能性が高い。ですから、全ての格子点のスピン $\sigma$ が一致します。これは、スピンを各格子点にある微小な磁石の向きだと思うと、温度が非常に低いときは系全体が自発的に一方向に磁化している、と思えます。一方、$T=\infty$ では、各格子点は独立ですから、系全体としては完全に乱雑になっています。これは高温では磁石の磁化が消えた、と思う事ができます。$T$ を $0$ から $\infty$ まで徐々に変化させた場合、一体どこで磁化が消えるのだろうか? 磁化が消える瞬間はどのようになるのか、ということを考えましょう。

\begin{figure}
\[
\begin{array}{ccc}
\includegraphics[width=.2\textwidth]{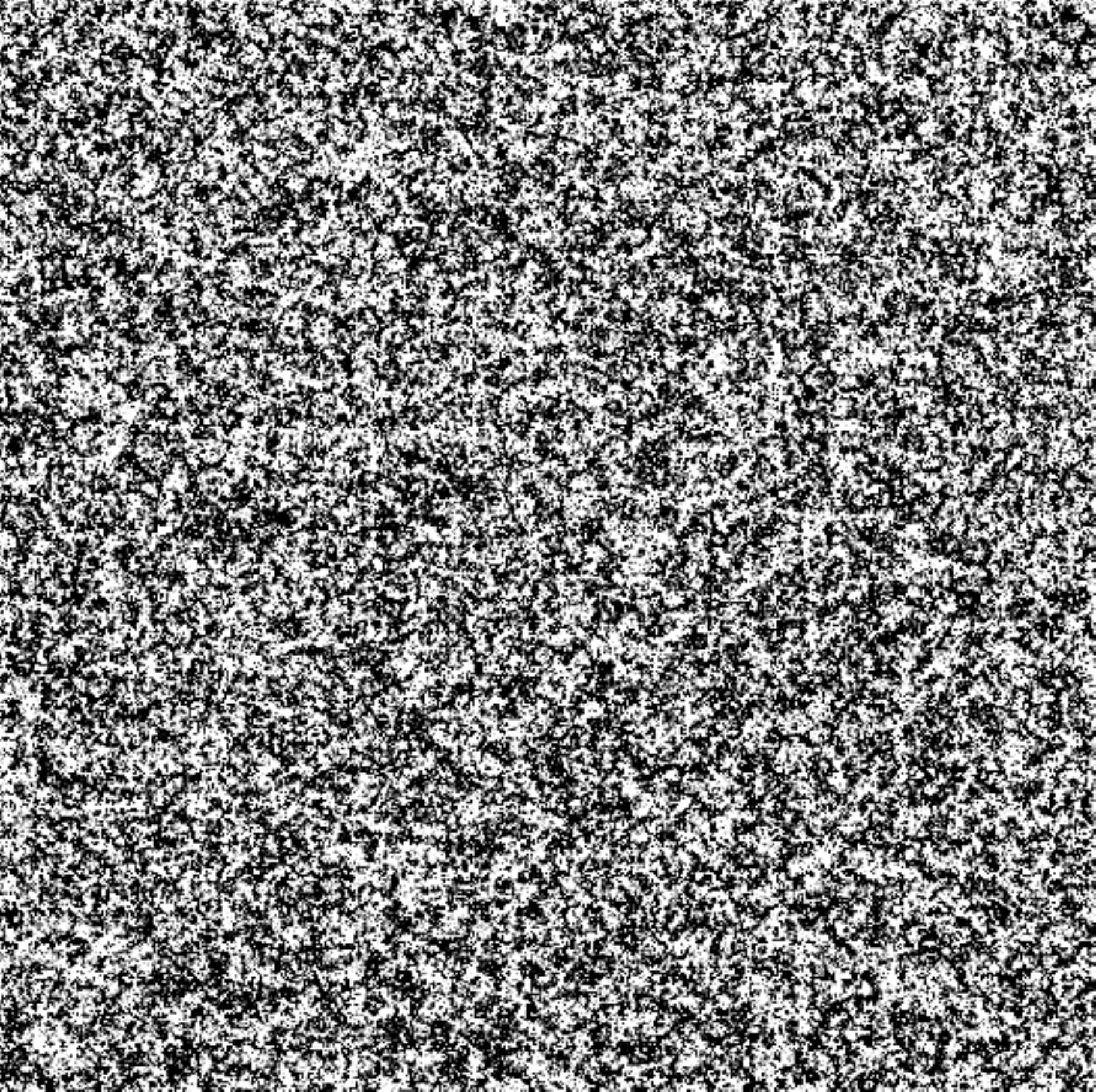} &
\includegraphics[width=.2\textwidth]{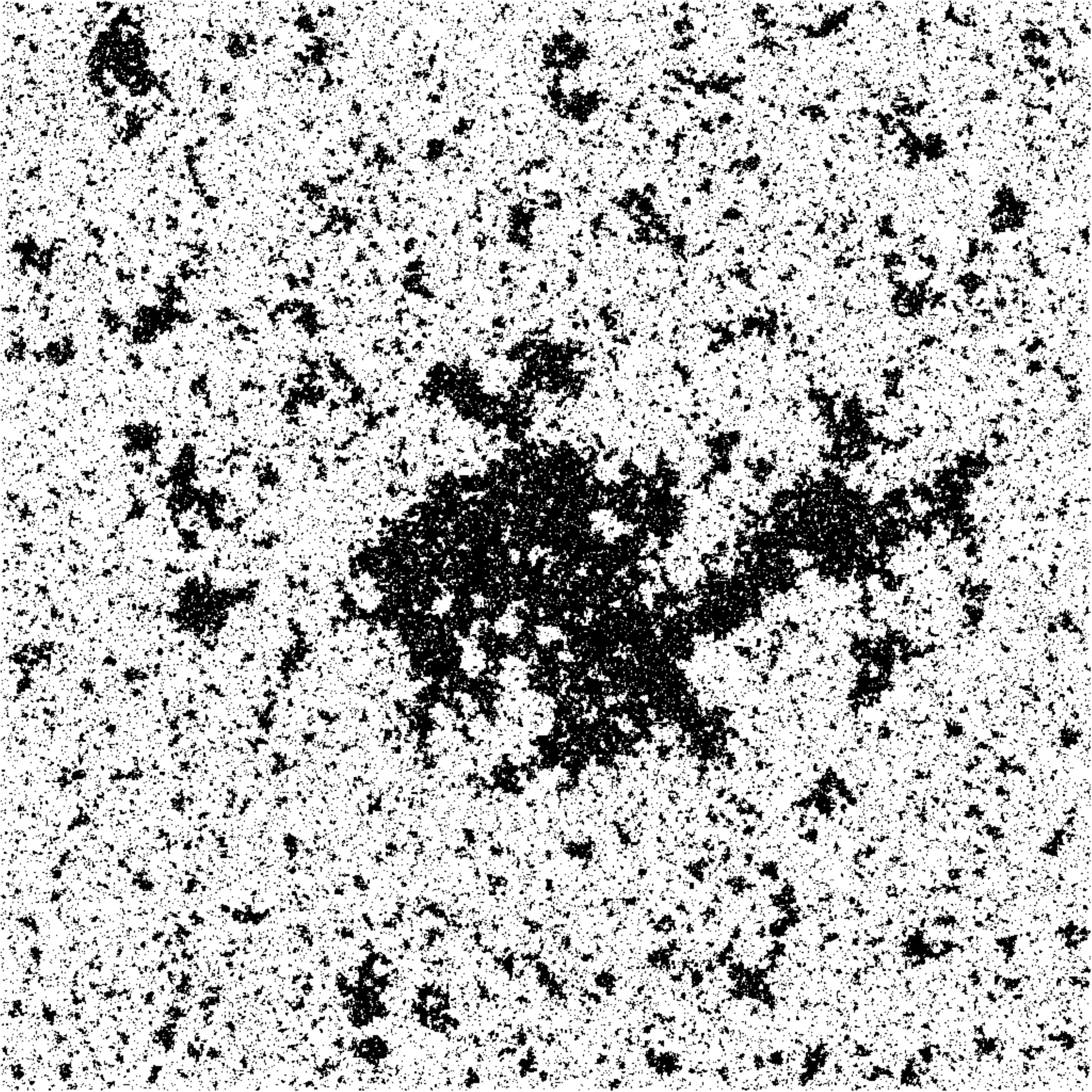} &
\includegraphics[width=.2\textwidth]{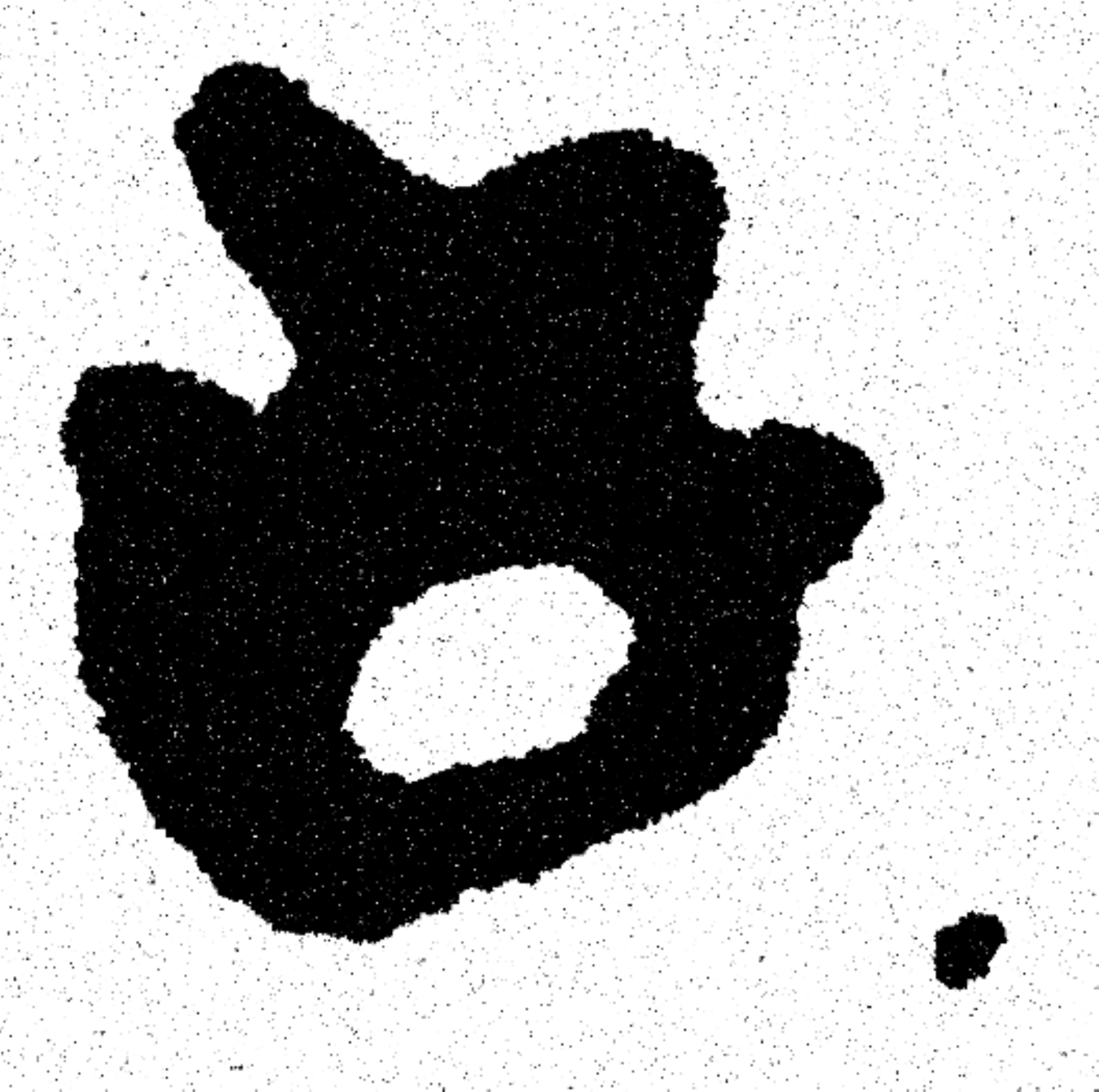}  \\
\text{高温相} &
\text{臨界} &
\text{低温相} 
\end{array}
\]
\caption{イジング模型の温度依存性\label{ising}}

\[
\vcenter{\hbox{\includegraphics[width=.2\textwidth]{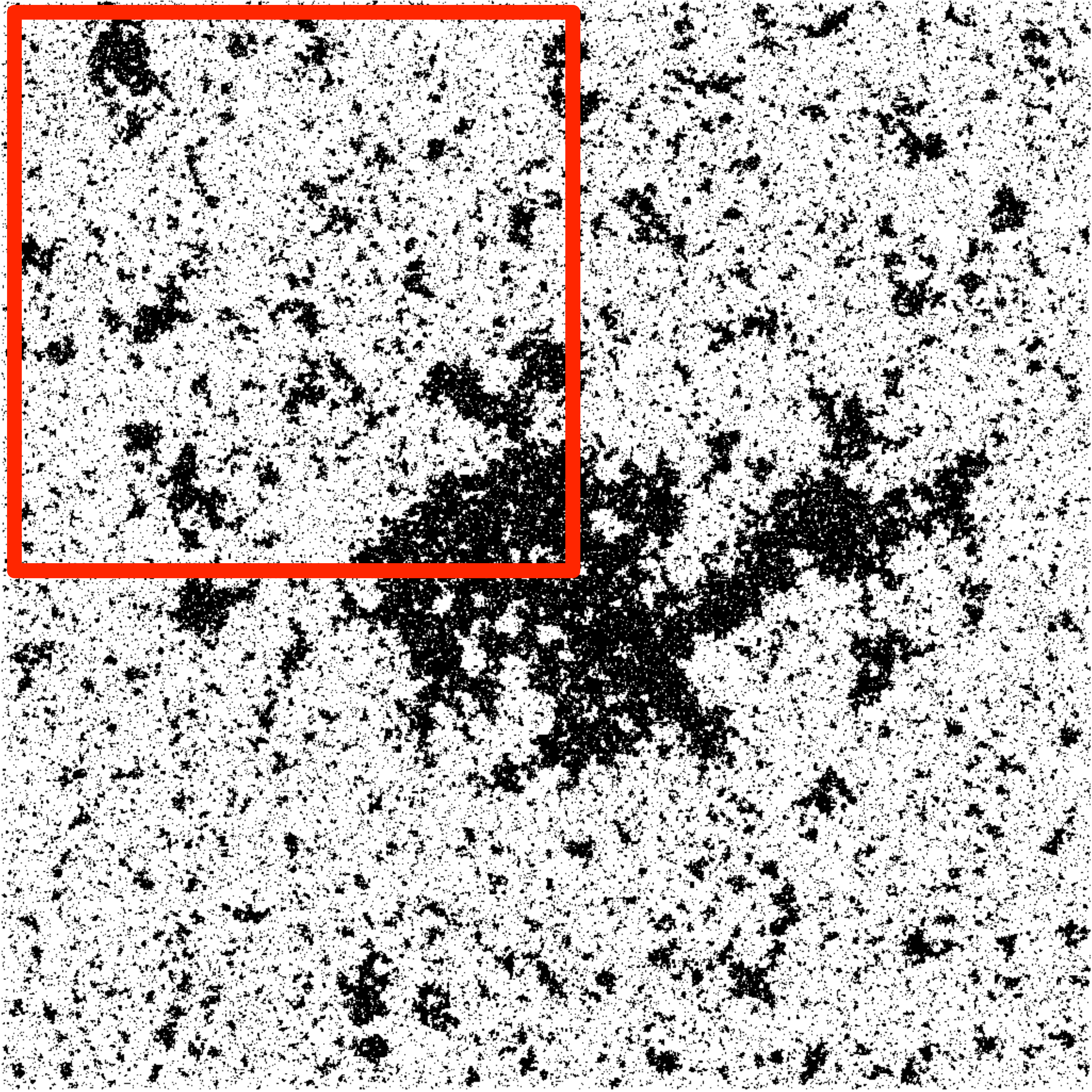}}}
\longrightarrow
\vcenter{\hbox{\includegraphics[width=.2\textwidth]{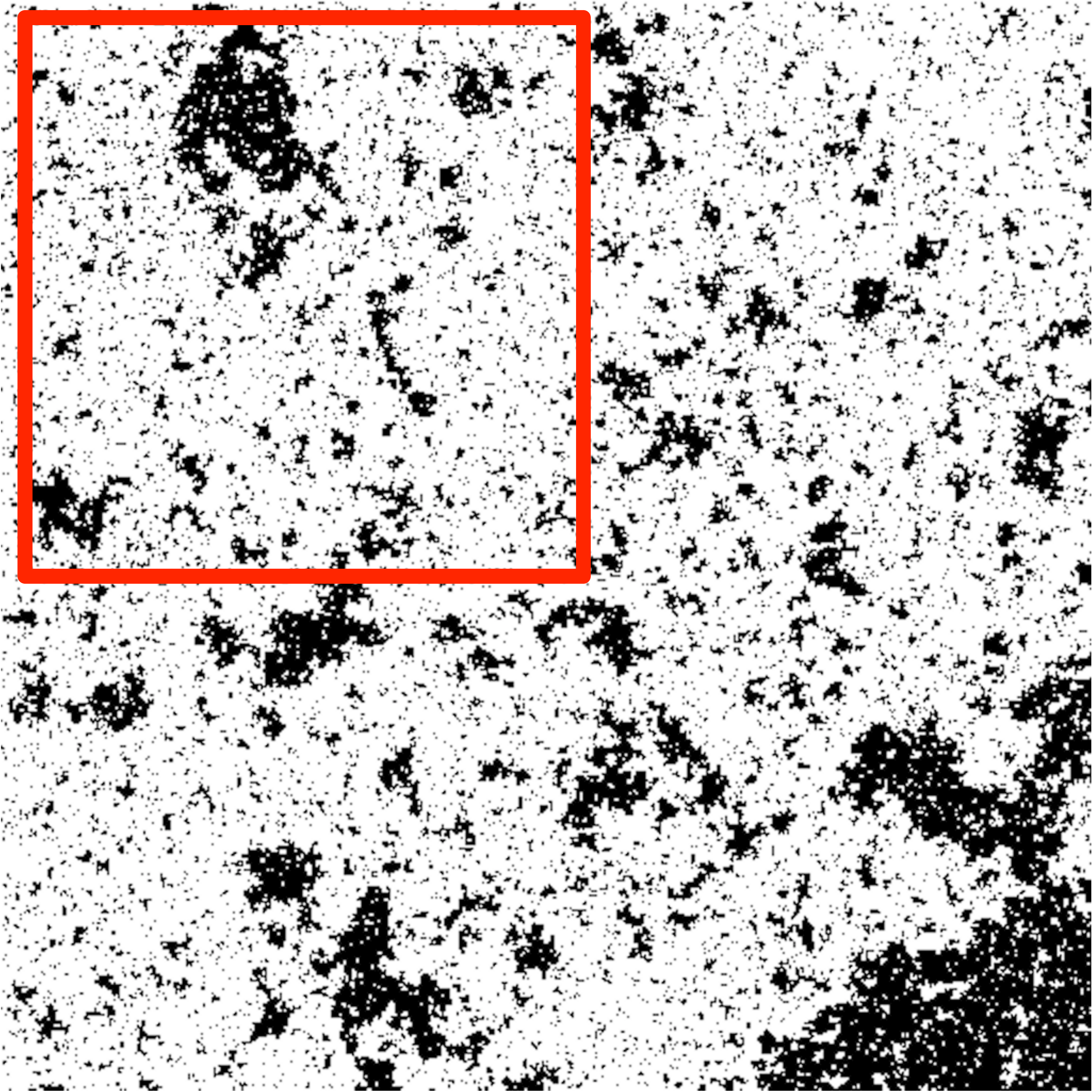}}}
\longrightarrow
\vcenter{\hbox{\includegraphics[width=.2\textwidth]{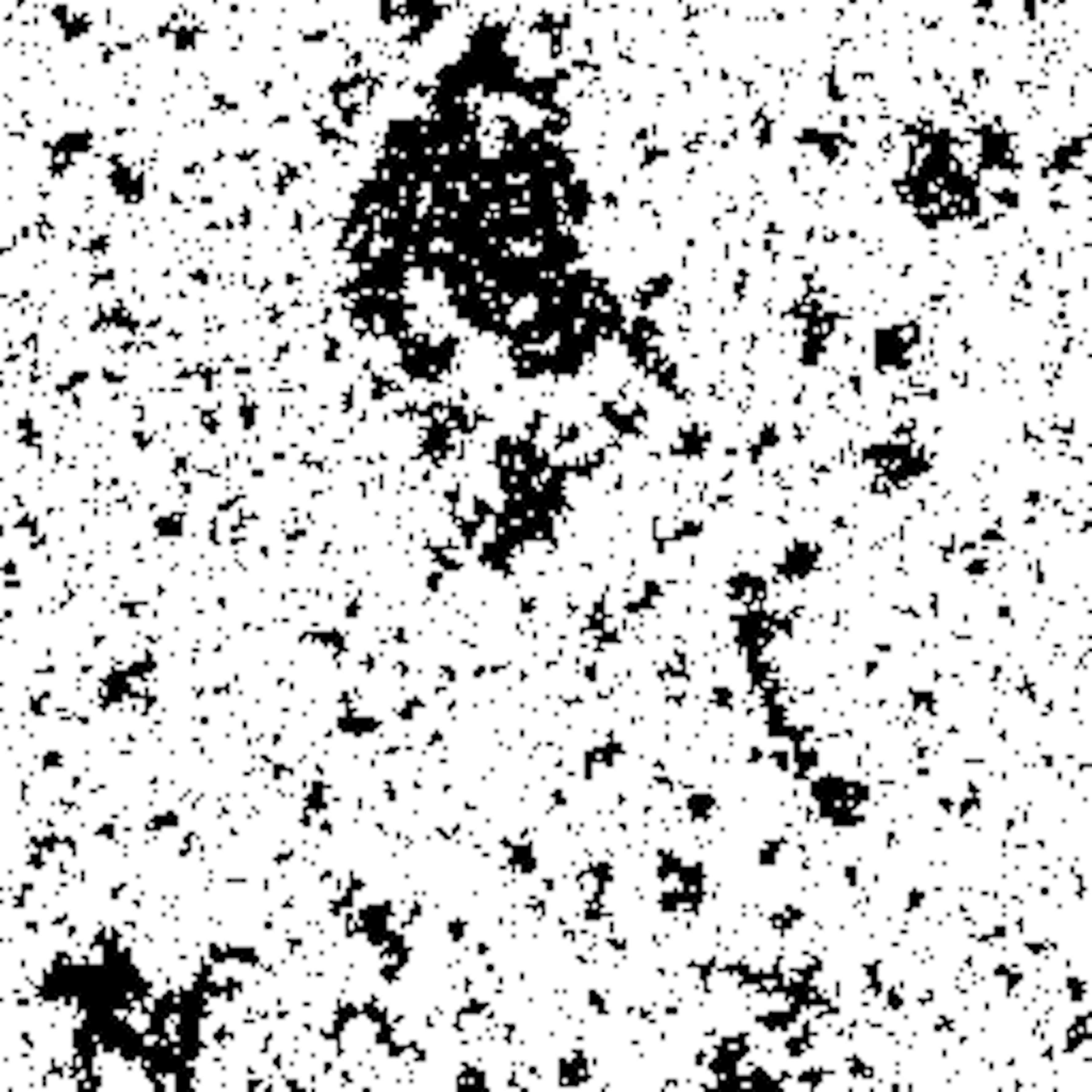}}}
\]
\caption{臨界温度でのイジング模型\label{scaling}}
\end{figure}

この模型は厳密に解かれているので、それを解説することも出来ますが、最近はパソコンが非常に強力なので、スピン配位をこの確率に従って生成して、何が起こっているかをみるのも悪くありません。しばらく遊ぶと、$T=2.2$ あたりで磁化が消滅することがわかります (図\ref{ising})。厳密解では、\begin{equation}
T_c=\frac{2}{\log(1+\sqrt{2})}
\end{equation}となることが知られています。さて、丁度温度をこの臨界点 $T_c$ にすると、スピンの揃っている塊の大きさにいろいろなものがあることが判ります。ひとつ配位を取って、二倍、二倍と拡大してみると、だいたい見た目が一定であることが判ります (図\ref{scaling})。もっと定量的には、$\vev{\sigma_{0,0} \sigma_{x,0}}$ を測定すると、両対数グラフで直線上にのることが判ります (図\ref{spinspin})。厳密解では \begin{equation}
\vev{\sigma(x)\sigma(0)}\propto 1/x^{1/4}\hbox{。}
\end{equation} ですから、スケールの変更 
$\vec x\to \alpha \vec x$ に伴って 
\begin{equation}
\sigma \to \alpha^{1/8} \sigma \label{spintr}
\end{equation} としてやれば系は不変です。
これを、系のスケール変換 $\vec x\to \alpha\vec x$ のもとでの不変性と言います。

\begin{figure}
\[
\includegraphics[width=.4\textwidth]{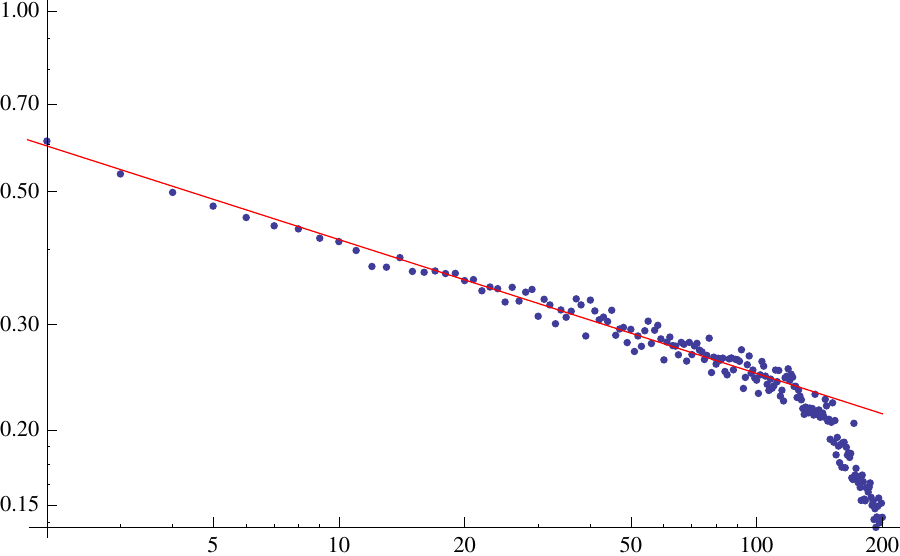}
\]
\caption{臨界温度での $\vev{\sigma_{0,0} \sigma_{x,0}}$ の振舞。\label{spinspin}}
\end{figure}

\begin{figure}\[
\begin{array}{cc}
\vcenter{\hbox{\includegraphics[width=.2\textwidth]{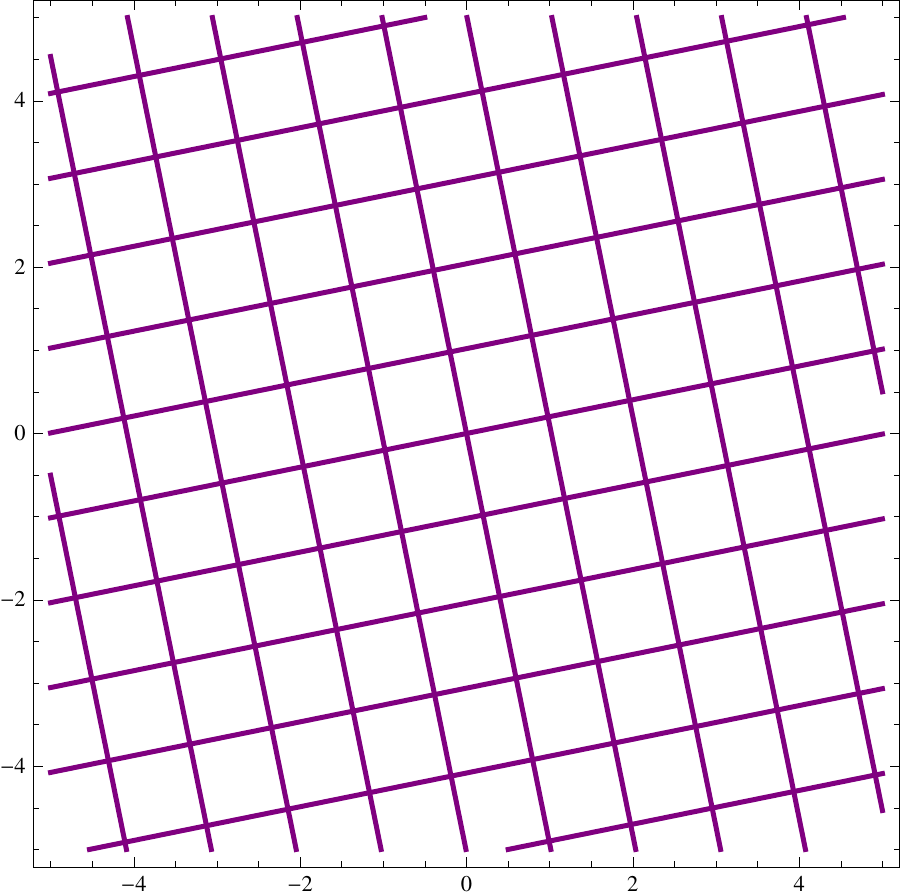}}}&
\vcenter{\hbox{\includegraphics[width=.2\textwidth]{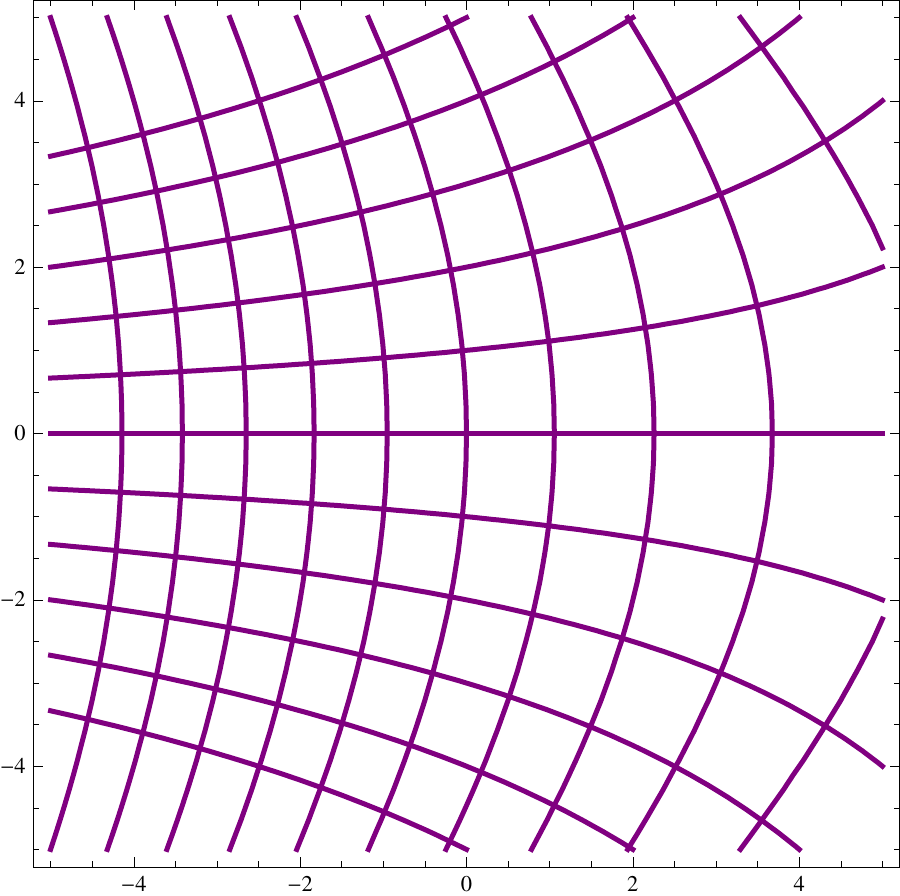}}} \\
z\mapsto e^{i\theta}\cdot z & 
z\mapsto z+ a z^2
\end{array}
\]
\caption{共形変換の例\label{conformal}}
\end{figure}

ある緩い条件を満たす二次元系では、もし系がスケール変換で不変ならば、自動的にもっと一般の局所的な角度を保つ変換のもとで理論が不変になることが知られています。二次元の座標を $z=x+iy$ と書くと、一般に正則関数 $f(z)$ を用いて \begin{equation}
z \mapsto z'=f(z)
\end{equation} という操作が局所的な角度を保ちます (図\ref{conformal})。微小変換は \begin{equation}
z \mapsto z'=z+ \sum_n \epsilon_n z^{n+1}
\end{equation} ですから、無限個のベクトル場 $\xi_n = z^{n+1} \partial_z $ で系が不変であることになります。 複素共役の $\bar\xi_n = \bar z^{n+1}\bar\partial_{\bar z}$ もあります。これらの交換関係は、計算すると \begin{equation}
[\xi_m,\xi_n] = (m-n) \xi_{m+n}\hbox{、} \quad
[\bar\xi_m,\bar\xi_n] = (m-n) \bar\xi_{m+n}\hbox{、} \quad
[\xi_m,\bar\xi_n]=0
 \label{Witt}
\end{equation} となります。

\begin{figure}
\[
\includegraphics[width=.3\textwidth]{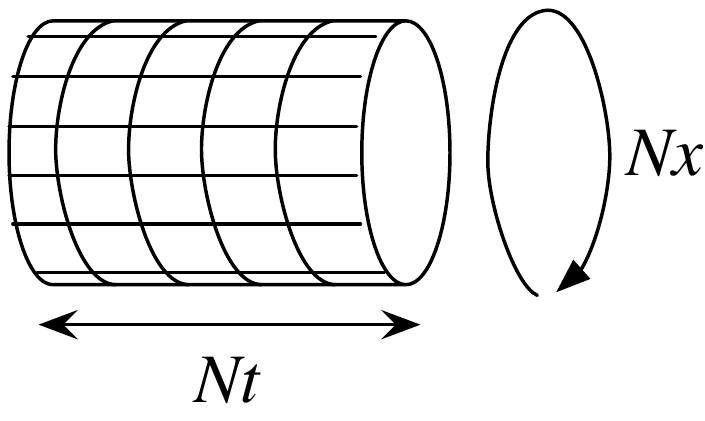}
\]
\caption{円柱上のイジング模型\label{ising-on-cyl}}
\end{figure}

さて、すこし見方を変えて、この二次元系のうち $x$ 方向は空間で、$y$ 方向は時間だと思いましょう: $t=y$。
無限に広いとややこしいので、$x$ 方向は $N_x$ 個、$t$ 方向は $N_t$ 個格子点があるとします(図\ref{ising-on-cyl})。
ある時刻 $t$ を固定すると、スピン $\sigma_i$ ($-\infty<i<\infty$) がありますが、一つ一つの配位に対してベクトル空間の基底 $\ket{\sigma_x}$ を考えましょう。この空間 $\cH$ には合計 $2^{N_x}$ 個基底があります。$\cH$ に作用する行列を二つ考えます: \begin{align}
A\ket{\sigma_x} &= \exp\left[\frac1T\sum_{x}\sigma_x\sigma_{x+1} \right] \ket{\sigma_x}\hbox{、} \\
B\ket{\sigma_x} &= \sum_{(\sigma'_x)}\exp\left[\frac1T\sum_{x}\sigma_x\sigma'_{x} \right] \ket{\sigma'_x}
\end{align} すると、分配関数 \eqref{isingZ}  は \begin{equation}
Z=\tr_\cH (AB)^{N_y} = \tr_\cH e^{-N_y H} \quad \text{但し}\quad H=\log (AB)
\end{equation} となり、時間 $N_t$ だけ量子力学的ハミルトニアン $H$ で系が発展したと思うことが出来ます。
いいかえれば、時間方向に動かすベクトル場 $\partial_t$ が、$\cH$ に作用する行列 $H$ に変わったわけです。

$N_x$ を非常に大きく取っておいて、この書き換えを臨界点 $T=T_c$ のときに行うと、時間並進だけでなく、共形変換 $\xi_n$、 $\bar\xi_n$ 全体が $\cH$ に作用する行列になります。それを $L_n$、 $\bar L_n$ と書きましょう。一般に、古典的なベクトル場を量子力学的な行列に焼き直すと、補正が入り得ます。無限個ある $\xi_n$ の交換関係 \eqref{Witt} を矛盾無く変更するには、次のようにするしかないと知られています: \begin{equation}
[L_m,L_n] = (m-n)L_{m+n} + \frac c{12}(m^3-m) \delta_{n,-m}\label{virasoro}
\end{equation}但し $-\infty<n,m<\infty$ は整数で $c$ は正の定数。これがビラソロ代数です。
$\bar\xi_n$ も全て演算子 $\bar L_n$ になり、同じ交換関係を満たします。
$L_n$ のエルミート共役は $L_{-n}=L_n^\dagger$ とします。以下簡単のため通常 $\bar L_{n}$ は忘れることにします。

\subsection{ビラソロ代数の表現}
ビラソロ代数の表現を調べる前に、調和振動子の量子化を復習しましょう。
運動量演算子 $p$ と位置演算子 $q$ が $[q,p]=i$ という交換関係を満たす際に、ハミルトニアン $H=(p^2+q^2)/2$ を調べたい。勿論 $p=\partial/\partial q $ として、微分演算子 $\partial^2/\partial q^2 + q^2$ の固有関数を特殊関数として調べても良いですが、代数的に考えましょう。そのために $a=p+iq$、$a^\dagger = p-iq$ を定義すると、\begin{equation}
H= a^\dagger a + \frac12 
\end{equation} と書き直せます。$[a,a^\dagger]=-1$ ですから、$[H,a]=-a$ です。さて、$H$ に固有値 $E$ の固有状態 $\ket{E}$ があったとしましょう: $H\ket E=E \ket E$。すると、\begin{equation}
H a \ket E = (a H - a) \ket E = (E-1) a\ket E\hbox{、}
\end{equation} すなわち、$a\ket E$ は 固有値 $E-1$ の固有ベクトルです。
この操作は何度でも繰り返せます。
一方で、状態空間が正値、すなわち勝手な状態 $\ket\psi$ に対して $\| \ket \psi \|^2 = \vev{\psi|\psi}\ge 0$ とすると、
\begin{equation}
E \| \ket E \|^2 =  \vev{E |  H | E} = \|  p \ket E \|^2 + \|q \ket E \|^2 \ge 0 
\end{equation}ですから、固有値 $E$ は非負です。よって、$a^n\ket E$ はいずれ消滅しないといけない、すなわち何か状態 $\ket{\text{vac}}$ があって \begin{equation}
a \ket{\text{vac}} = 0\hbox{。} 
\end{equation} すると \begin{equation}
 H \ket{\text{vac}} = \frac12 \ket{\text{vac}}\hbox{。}
\end{equation} また、同様に固有値 $E$ の状態に $a^\dagger$ を掛けると固有値 $E+1$ の状態になるのも示せます。よって、\begin{equation}
\ket{n+\frac12} = (a^\dagger)^n \ket{\text{vac}} 
\end{equation} が固有値 $\frac12+n$ の固有状態になります。

ですから、ハミルトニアン $H$ に対して、生成演算子 $a^\dagger$ はエネルギーを $1$ あげ、消滅演算子 $a$ はエネルギーを $1$ 下げ、消滅演算子 $a$ で消される状態が最低エネルギー状態 $\ket{\text{vac}}$ です。

ビラソロ代数の場合は、$L_0$ をハミルトニアンと思うのが都合が良いです: $H=L_0$。すると、\begin{equation}
[L_0,L_n] = -n L_n 
\end{equation} ですから、$L_{-n}$ がエネルギーを $n$ あげる生成演算子で、$L_{n}=L_{-n}^\dagger$  がエネルギーを $n$ 下げる消滅演算子です。全ての消滅演算子で消される状態が最低エネルギーですが、そこでの $H=L_0$ の固有値を $\Delta$ としましょう: \begin{equation}
L_{0} \ket{\Delta}=\Delta \ket{\Delta}\hbox{、} \qquad
L_{n} \ket{\Delta}=0 \qquad (n>0)\hbox{。}
\end{equation} 調和振動子の場合と異なり、$\Delta$ はこれだけでは定まりません。

\begin{figure}
\[
\includegraphics[width=.8\textwidth]{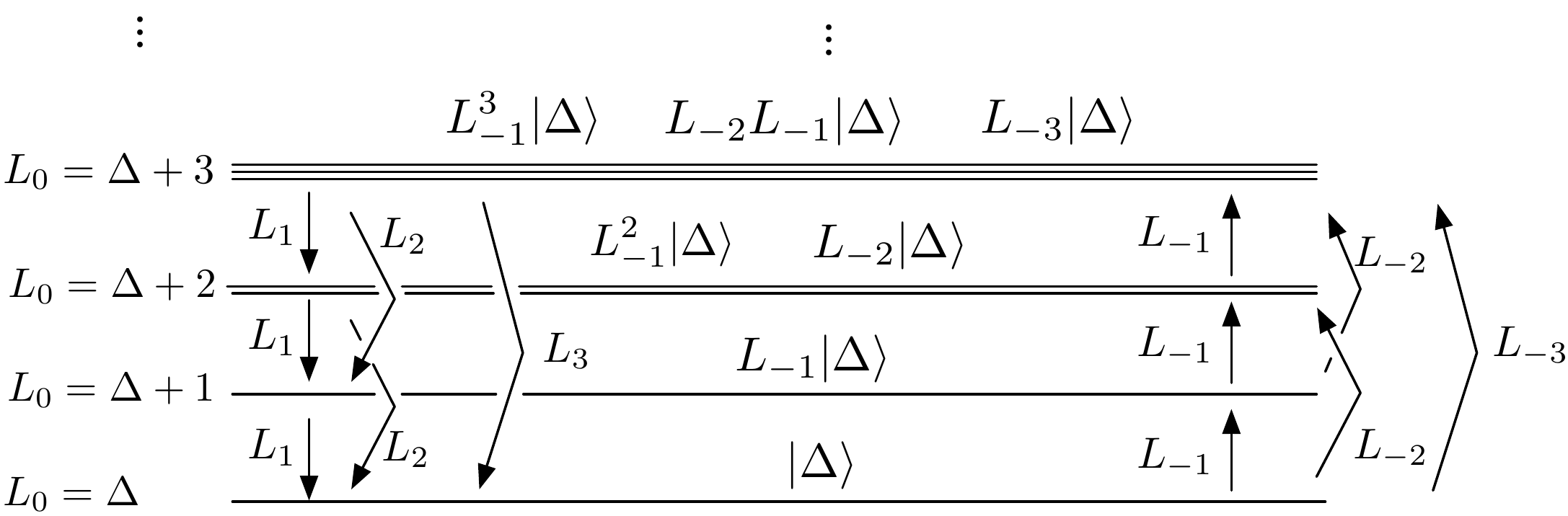}
\]
\caption{ビラソロ代数の表現 \label{verma}}
\end{figure}

一般に、$L_0$ の固有値が $\Delta+N$ である状態は $L_{-n_1}L_{-n_2}\cdots L_{-n_k}\ket\Delta$ で $\sum n_i=N$ となるようなものです(図\ref{verma})。
これを簡略に \begin{equation}
\ket{\Delta;N;n_1,\ldots,n_k} =L_{-n_1}L_{-n_2}\cdots L_{-n_k}\ket\Delta
\end{equation} と書きましょう、但し $n_1\ge n_2 \ge \cdots \ge n_k$ としておきます。
これらが全て線形独立な場合、この表現をビラソロ代数の Verma 表現と言います。

$N$ は通常次数 (grade) と呼ばれます。次数が 1 の状態は $L_{-1}\ket{\Delta}$ のみです。これのノルムは、交換関係をつかうと \begin{equation}
\vev{\Delta|L_{1}L_{-1}|\Delta}=2\Delta \vev{\Delta|\Delta}
\end{equation} となりますから、$\Delta$ は正です。

次に次数が 2 の状態は一般に \begin{equation}
\ket{\psi}=c_{11} \ket{\Delta;2;1,1} + c_2 \ket{\Delta;2;2} 
\end{equation}と書けますが、これのノルムは交換関係 \eqref{virasoro} をつかって計算すると \begin{equation}
\vev{\psi|\psi} = (\bar c_{11},\bar c_{2}) M_{N=2}  \begin{pmatrix}
c_{11} \\ c_2
\end{pmatrix}\quad\text{但し}\quad
M_2= \begin{pmatrix}
4\Delta(2\Delta+1) & 6\Delta \\
6\Delta & 4\Delta+c/2
\end{pmatrix}
\end{equation} となります。これが負にならないためには、\begin{equation}
4\Delta(2\Delta+1)(4\Delta+c/2) \ge (6\Delta)^2
\end{equation} でなければいけません。書き換えると、\begin{equation}
\Delta(\Delta-\Delta_{1,2})(\Delta-\Delta_{2,1})\ge 0
\end{equation} です、ただし \begin{equation}
\Delta_{r,s}=\frac{c-1}{24}+\frac14(r\alpha_++s\alpha_-)^2\hbox{、}\qquad \alpha_\pm=\frac{\sqrt{1-c}\pm\sqrt{25-c}}{\sqrt{24}}\hbox{。}
\end{equation}
$c<1$ ですと $\Delta_{1,2}$、 $\Delta_{2,1}$ は実ですから、$\Delta_{1,2}< \Delta < \Delta_{2,1}$ だと駄目なわけです。

一般に次数 $N$ のところにある状態の数は、$N$ を正の整数の和として書く場合の数だけあります。それを $p_N$ と書く事にしますと、 $M_{N}$ は $p_N\times p_N$ 行列になります。状態空間が正定値であるためには、全ての $N$ に対して $\det M_N\ge 0$ でないといけません。この行列式は Kac によって計算されており、\begin{equation}
\det M_N \propto \prod_{r,s\ge 1;\ rs\le N} (\Delta-\Delta_{r,s})^{p_{N-rs}}
\end{equation} となります。$c\ge 1$ のときはほぼ自動的にこれは正になりますが、$c<1$ のときは $\Delta$ が $\Delta_{r,s}$ のどれかでない限り、いずれ何らかの $\Delta_{r,s}$、 $\Delta_{s,r}$ に挟まれて駄目になってしまいます。
これらの条件を丁寧に調べると、状態空間が正定値であるためには、\begin{equation}
c\ge 1 \quad \text{もしくは $m\ge2$ なる正の整数を取って} c=1-\frac{6}{m(m+1)} 
\end{equation}となることが知られています。さらに、後者の場合は $\Delta$ の値は 
$r,s$ を $1\le s \le r <m$ なる正の整数として$\Delta = \Delta_{r,s}$
に限られます。\begin{equation}
\alpha_+=\frac{m+1}{\sqrt{m(m+1)}}\hbox{、}\quad
\alpha_-=\frac{-m}{\sqrt{m(m+1)}}
\end{equation}ですから、
\begin{equation}
\Delta_{r,s}=\frac{((m+1)r-ms)^2-1}{4m(m+1)} 
\end{equation} となります。
$m=2$ のときは $c=0$、 許される $\Delta$ は $\Delta=0$ のみで面白くありません。
次の $m=3$ の場合は、$c=1/2$、 許される $\Delta$ は 
\begin{equation}
\Delta_{1,1}=0\hbox{、}  \quad  \Delta_{2,2}=\frac1{16}\hbox{、} \quad  \Delta_{1,2}=\frac12 \label{isingdims}
\end{equation}の三種類です。

イジング模型は臨界点では丁度この $c=1/2$ のビラソロ代数が $\xi_n$ から来る $L_n$ と $\bar\xi_n$ から来る $\bar L_n$ とがあります。
スケール変換 $(x,y)\to \alpha(x,y)$ は微小変換 $\alpha=1+\epsilon$ に対しては $z=x+iy$ で書いて \begin{equation}
z\frac{\partial}{\partial z}
+\bar z\frac{\partial}{\partial \bar z}  = \xi_0+\bar \xi_0
\end{equation} で与えられます。すなわち $L_0+\bar L_0$ です。
スピン演算子の変換性を \eqref{spintr} に書きましたが、微小変換を考えると $L_0+\bar L_0$ の固有値が $1/8$ であることになります。これは \eqref{isingdims} で $L_0$ 及び $\bar L_0$ の固有値が両方とも $1/16$ になっていることに対応します。

\subsection{ビラソロ代数のコヒーレント状態}
さて、話をまた調和振動子に戻して、コヒーレント状態を考えましょう。
$p$ と $q$ は交換しませんから、同時固有状態をとることはできません。
ある状態 $\ket\psi$ に対して、$\vev{O}=\vev{\psi|O|\psi}$ と略記することにすると、$p$ と $q$ の広がりは \begin{equation}
(\delta p)^2=\vev{(p-\vev{p})^2}\hbox{、} \qquad
(\delta q)^2=\vev{(q-\vev{q})^2}
\end{equation} と思えますから、\begin{equation}
\delta p^2 \delta q^2 = \| (p-\vev{p}) \ket\psi \|^2 \| (q-\vev{q}) \ket\psi \|^2 \ge \left[\mathrm{Im} \vev{\psi| (p-\vev{p})(q-\vev{q}) |\psi} \right]^2= \frac14
\end{equation} となるのでした。不確定性を最小にする状態は可能な限り古典的な状態と言っても良いでしょう。上の式変形を等式にする一つの方法は \begin{equation}
i(p-\vev{p})\ket\psi 
= 
(q-\vev{q})\ket\psi 
\end{equation} とすればよいです。$\lambda=\vev{p+iq}$ とすると、$\psi$ が消滅演算子 $a=p+iq$ の固有状態であることがわかります:
\begin{equation}
a\ket{\psi}=\lambda\ket{\psi}\hbox{。}
\end{equation}  これをコヒーレント状態と呼ぶのでした。以下固有値 $\lambda$ のコヒーレント状態を $\ket{\lambda}$ と書く事にしましょう。

調和振動子の固有状態はコヒーレント状態の一例です: $\ket{\text{vac}}=\ket{0}$。
$[a,a^\dagger]=1$ であることを利用して、\begin{equation}
a e^{\lambda a^\dagger} \ket{0} = \lambda e^{\lambda a^\dagger} \ket{0} 
\end{equation} すなわち $\ket{\lambda}=e^{\lambda a^\dagger}\ket 0$ です。状態を規格化するには、\begin{equation}
\vev{\lambda|\lambda}=\vev{0| e^{\bar\lambda a}e^{\lambda a^\dagger} |0} = e^{\bar\lambda\lambda} \vev{0|0}
\end{equation} とすればよいです。

調和振動子のコヒーレント状態はいろいろな応用があります。ビラソロ代数も重要です。ですから、ビラソロ代数のコヒーレント状態を考える事も意味が無くはないでしょう。
$\ket{\Delta}$ で生成される Verma 表現の中のベクトル $\ket\psi$ で、消滅演算子 $L_n$ ($n>0$) が固有値を持つものを考えます: \begin{equation}
L_n \ket\psi = \lambda_n \ket\psi\hbox{。}
\end{equation} 交換関係から、すぐに $n\ge 3$ なら $\lambda_n=0$ とわかります。簡単のために $\lambda_2$ もゼロとしてしまって、$\lambda\equiv \lambda_1$ で指定される状態 $\ket{\Delta,\lambda}$ を考えましょう: \begin{equation}
L_1\ket{\Delta,\lambda}= \lambda \ket{\Delta,\lambda}, \quad L_2\ket{\Delta,\lambda}=0\hbox{。} 
\end{equation} すると  $n>2$ について $L_n\ket{\Delta,\lambda}=0$ は自動的に従います。
調和振動子にならって、$e^{\lambda L_{-1}}\ket{\Delta}$ を考えたいところですが、ビラソロ代数の交換関係はそれほど簡単でないため、それでは安直すぎます。

まず愚直に計算してみましょう。欲しい状態は、なんにせよ展開できる筈ですから、\begin{equation}
\ket{\Delta,\lambda}=\ket{\Delta} + c_1 L_{-1} \ket{\Delta} + c_{11} L_{-1}^2 \ket{\Delta} + c_2 L_{-2} \ket{\Delta} + \cdots
\end{equation}と書きます。すると、  \begin{equation}
L_1 c_1  L_{-1} \ket{\Delta}= \lambda \ket{\Delta}
\end{equation}  から $c_1= \lambda/(2\Delta)$ と定まり、\begin{align}
L_1 (c_{11} L_{-1} ^2 \ket{\Delta} + c_2 L_{-2} \ket{\Delta})&= \lambda c_1 L_{-1} \ket{\Delta}, &
L_2 (c_{11} L_{-1} ^2 \ket{\Delta} + c_2 L_{-2} \ket{\Delta})&=0
\end{align} から $c_{11}$、 $c_{2}$ が定まります。長さの二乗は、これより \begin{equation}
\vev{\Delta,\lambda|\Delta,\lambda}=1+\frac{\lambda^2}{2\Delta}+\frac{\lambda^4(c+8\Delta)}{4\Delta((1+\Delta)c-10\Delta+16\Delta^2)}+\cdots
\end{equation} となります。

さて、この調子で下から係数はすべて決められるのでしょうか? 次数 2 では、未知数が $c_{11}$ と $c_{2}$ が丁度二つ、方程式は次数 1 に状態が 1 つ、次数 0 にも状態が 1 つあったので、こちらも 2 つで無事解くことができました。これは $p_2=p_1+p_0$ だったという偶然に基づくもので、一般には $N$ が大きくなると $p_N \ll p_{N-1}+p_{N-2}$ となり、方程式の数が過剰になって解くのが一見困難になります。実際に計算をしてみると、それでも無事に解く事ができることがわかります。

その理由はこうです\cite{Marshakov:2009gn}。欲しいコヒーレント状態があったとして、それを次数毎にまとめて \begin{equation}
\ket{\Delta,\lambda}=\ket{\psi_0}+\lambda\ket{\psi_1}+\lambda^2\ket{\psi_2}+\cdots 
\end{equation} とします。すると、 \begin{equation}
L_1\ket{\psi_N} = \ket{\psi_{N-1}}\hbox{、} \quad
L_2\ket{\psi_N}=0
\end{equation} となっているはずです。すると、簡単にわかるように、\begin{equation}
\vev{\Delta| (L_1)^N |\psi_N}=1\hbox{、}
\end{equation} また、それ以外の組み合わせでは \begin{equation}
\vev{\Delta| (L_1)^{N-2} L_2 |\psi_N} =0
\end{equation} 等となります。すなわち、$\ket{\psi_N}$ は、次数 $N$ の状態の中で、 $L_{-1}^N \ket{\Delta}$ とのみ内積を持って、他の基底とは直交しているわけです。そこで、\begin{equation}
\ket{\psi_N}=\sum_{i_1+\cdots+i_k=N} (M^{-1}_N)^{11\cdots1,i_1i_2\cdots i_k} L_{-i_1}L_{-i_2} \cdots L_{-i_k} \ket{\Delta} 
\end{equation} と取ればよいことがわかりました。
これから、$\ket{\Delta,\lambda}$ の内積もすぐ計算でき、\begin{equation}
\vev{\Delta,\lambda | \Delta,\lambda}= 1 + \lambda^2 (M^{-1}_1)^{1,1}+ \lambda^4 (M^{-1}_2)^{11,11}+ \lambda^6 (M^{-1}_3)^{111,111} +\cdots\label{russian}
\end{equation} となります。
すぐわかるように $e^{tL_0} \ket{\Delta,\lambda}= e^{t\Delta} \ket{\Delta,e^t \lambda}$ ですから、\begin{equation}
\vev{\Delta,\lambda | \Delta,\lambda}= \vev{\Delta,1 | \lambda^{2(L_0-\Delta)} | \Delta,1} 
\end{equation} と書き直せます。 $L_0$ が円柱に二次元共形場理論を置いたときの時間発展の演算子だったことを思い出しますと、この節で計算した量は、図 \ref{HH} のように図示することができます。

\begin{figure}\[
\includegraphics[width=.3\textwidth]{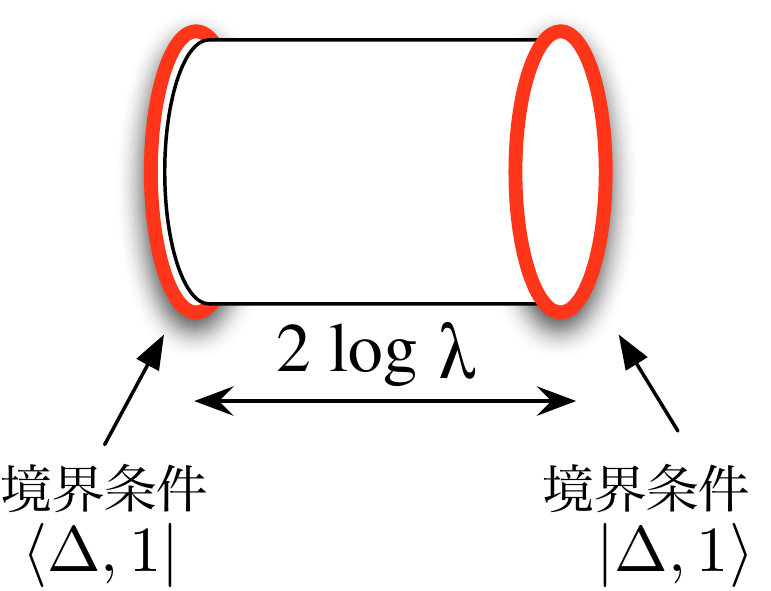}
\]
\caption{量 $\vev{\Delta,\lambda | \Delta,\lambda}$ は、長さ $2\log\lambda$ の円柱の両側に状態 $\ket{\Delta,1}$ が境界条件として与えられているものと思える。 \label{HH}}
\end{figure}

\section{四次元ゲージ理論とインスタントンの統計力学}
\subsection{非可換ゲージ理論}
さて、すっかり話を変えて、この節では四次元のゲージ理論の話をします。(この節の内容の詳細は、例えば教科書 \cite{Coleman} や、講義録\cite{tHooft:1999au}を参照のこと。)
ゲージ理論の一番簡単な例は Maxwell の電磁気学です。電場 $\vec E$ と磁場 $\vec B$ は相対論的形式では $F_{\mu\nu}=-F_{\nu\mu}$ にまとまるのでした: \begin{equation}
F_{0i}=E_i\hbox{、}\quad 
F_{12}=B_3\hbox{、} \ F_{23}=B_1\hbox{、}\ F_{31}=B_2\hbox{。}
\end{equation}すると、方程式は \begin{equation}
\partial_\mu F_{\mu\nu}=0\hbox{、} \qquad \partial_{\mu} F_{\nu\rho} + \partial_{\nu} F_{\rho\mu} + \partial_{\rho} F_{\mu\nu}=0  \label{MaxwellEOM}
\end{equation}と書けます。但し前者では拡張された Einstein の規約、すなわち同じ添字が二度現れると適切に計量 $\eta^{\mu\nu}$ を入れて足し上げるという規約を使いました: \begin{equation}
\partial_\mu F_{\mu\nu}= \sum_{\rho,\mu}\partial_\rho \eta^{\rho\mu} F_{\mu\nu}\hbox{。}
\end{equation}
\eqref{MaxwellEOM}の後者は四元ベクトルポテンシャル $A_\mu$ を用いて \begin{equation}
F_{\mu\nu}=\partial_\mu A_\nu -\partial_\nu A_\mu 
\end{equation}とすると自動的に解くことが出来ます。但し、$F_{\mu\nu}$ から $A_\mu$ は一意的には定まらず、$\chi$ を勝手なスカラー関数として \begin{equation}
A_\mu \mapsto A_\mu+\partial_\mu \chi	\label{MaxwellGauge}
\end{equation} としても $F_{\mu\nu}$ は変わりません。これをゲージ変換と言います。
もうひとつ重要な点は、 $A_\mu$ を基本的な力学変数だと思うと作用を書く事ができます、すなわち \begin{equation}
S=\int d^4x \frac14 F_{\mu\nu} F_{\mu\nu}
\end{equation} を変分することによって \eqref{MaxwellEOM} が出て来ます。
Maxwell 方程式の簡単でうれしいところは、その線形性です。すなわち、$A_\mu$ と $A'_\mu$ を 2 つの解とすると、$A_\mu+A_\mu'$ も自動的に解になります。

以上は古典論でしたが、これを量子論にするには、 Feynman の経路積分をする必要があります、すなわち \begin{equation}
Z=\int [DA_\mu] e^{iS}
\end{equation} として、可能なすべてのベクトルポテンシャル $A_\mu$ の配位に対して、位相 $e^{iS}$ をつけて積分せよ、という操作です。
以上は計量が $\eta=\diag(-1,+1,+1,+1)$ のミンコフスキ空間での議論でしたが、以下簡単のため計量が $\delta=\diag(+1,+1,+1,+1)$ のユークリッド空間に話を変えることにします。すると経路積分は \begin{equation}
Z=\int [DA_\mu] e^{-S}
\end{equation}となって、未だ無限次元の積分ですが少しは扱いやすくなります。

20世紀後半の物理の大きな発見のひとつは、電磁気以外にあるこの世の他の2つの力、「強い力」と「弱い力」がどちらもこの Maxwell 理論の拡張である非可換ゲージ場の理論で書かれるということでした。
まず、非可換群 $\SU(N)$ をおさらいしましょう。$g\in \SU(N)$ は複素 $N\times N$ 行列で、ユニタリ $g^\dagger g=1$、 さらに $\det g=1$ としたものです。$\SU(2)$ は特に \begin{equation}
g = \begin{pmatrix}
z & -\bar w \\
w & \bar z
\end{pmatrix}
\end{equation} で且つ $|z|^2+|w|^2=1$ というものですから、$\SU(2)\simeq S^3$ であることがわかります。

$g$ が単位元に近いとして、 $g=1+\epsilon+\cdots $ と書きますと、$g^\dagger g=1$ から
$\epsilon+\epsilon^\dagger=0$、 更に $\det g=1$ から $\tr \epsilon=0$ となります。
というわけで、反エルミートでトレースがゼロの行列全体を $\SU(N)$ のリー代数と言います。

これをつかって、ゲージポテンシャル $A_\mu$ が $\mu=1,2,3,4$ に対して $\SU(N)$ のリー代数に入っているとしましょう。
ゲージ場の強さを \begin{equation}
F_{\mu\nu}=\partial_\mu A_\nu  -\partial_\nu A_\mu +[A_\mu,A_\nu]
\end{equation} と定めると、運動方程式は \begin{equation}
\partial_\mu F_{\mu\nu} + [A_\mu,F_{\mu\nu}]=0 \label{YMEOM}
\end{equation} となります。これが Yang-Mills 場の方程式です。これを与える作用は \begin{equation}
S=-\frac{1}{2 g^2} \int \tr F_{\mu\nu} F_{\mu\nu} 
\end{equation}です。実際に変分してみてください。$N\times N$ 行列の場合に、これを $\SU(N)$ ゲージ理論と呼びます。$g$ は結合定数と呼ばれます。

「弱い力」はこれで $N=2$ としたもの、「強い力」は $N=3$ として、さらに量子論にしたもので書かれることがわかっています。
実際に計算するには、やはり経路積分をします: \begin{equation}
Z=\int [DA_\mu] e^{-S} \label{YMpathintegral}
\end{equation} 勿論、これは無限次元積分になって、厳密な定義が出来ると Clay 賞の半分を取ったようなものですが、数学的厳密さにそれほどこだわらなければ、例えば時空 $\bR^4$ を非常に細かな格子 $\bZ^4$ で近似して、超巨大スーパーコンピューターで積分を計算してやることができます。そうすると、たとえばいろいろなハドロンの質量比が計算誤差の範囲できちんと再現できることがわかっています。

さて、Maxwell 場の作用がゲージ変換 \eqref{MaxwellGauge} で不変だったように、非可換ゲージ場の作用もゲージ変換で不変です。まず、$A_\mu$ に対するゲージ変換は時空から $\SU(N)$ への写像 $g(x)$ を使って \begin{equation}
A_\mu \mapsto g A_\mu g^{-1} + g \partial_\mu g^{-1}   \label{YMgauge}
\end{equation}とするものとします。すると、$F_{\mu\nu}$ に対しては \begin{equation}
F_{\mu\nu} \mapsto g F_{\mu\nu} g^{-1}
\end{equation} となり、作用の被積分関数は \begin{equation}
\tr F_{\mu\nu} F_{\mu\nu} \mapsto  \tr g F_{\mu\nu} g^{-1} g F_{\mu\nu} g^{-1} =\tr F_{\mu\nu} F_{\mu\nu}
\end{equation} となることが示せます。
ですから、通常 \begin{equation}
\lim_{|x|\to \infty} g(x) = 1
\end{equation}  となる $g(x)$ でのゲージ変換で結びつく二つの Yang-Mills 方程式の解は同一視します。これを、局所ゲージ変換での同一視と呼びましょう。
一方で、 $g(x)\equiv g$ と場所に依らない場合は \eqref{YMgauge} での微分項が落ちます。
これを大域ゲージ変換と呼び、大域ゲージ変換で結びつく二つの解は同一視しないことにします。

Yang-Mills 場の運動方程式 \eqref{YMEOM} には $A$ について三次の項がありますから、非線形です。解 $A_\mu$ と $A'_\mu$ とが二つ得られても、$A_\mu + A'_\mu$ は解にはなりません。ですから、フーリエ変換すれば解が求まるというわけでもありません。しかし、インスタントンと呼ばれる一連の具体的な解が知られており、それらは非常に詳細に調べる事ができます。そこで、それについて次に説明しましょう。

\subsection{インスタントン}
Yang-Mills 場の経路積分 \eqref{YMpathintegral} をすることを考えましょう。$S$ が小さいほうが寄与が大きいですから、$S$ を最小化する配位が重要そうです。
それを調べる為に、まず双対場 
\begin{equation}
\tilde F_{\mu\nu} = \epsilon_{\mu\nu\rho\sigma} F_{\mu\nu}/2
\end{equation} を導入しましょう。これは、非相対論的に成分を電場 $\vec E$ と磁場 $\vec B$ に分ければ、ちょうど $\vec E$ と $\vec B$ を入れ替えたものです。

勝手な配位 $A_\mu$ が与えられた際、作用が発散しないように無限遠で充分速く $F_{\mu\nu}\to 0$ となっているとします。このとき、$k$ を整数として \begin{equation}
-\int d^4 x \tr F_{\mu\nu} \tilde F_{\mu\nu} = -4 \int d^4 x \tr \vec E\cdot \vec B=16\pi^2 k 
\end{equation} となることが知られています。
ですから、$A_\mu$ の配位の空間は無限次元ですが、それは整数 $k$ でラベルされる部分空間にわかれているわけです (図\ref{components})。
これは下ですぐに示しますので、それまで認めて頂きましょう。この $k$ は物理ではインスタントン数、数学では第二チャーン数と呼ばれます。

\begin{figure}
\[
\includegraphics[width=.8\textwidth]{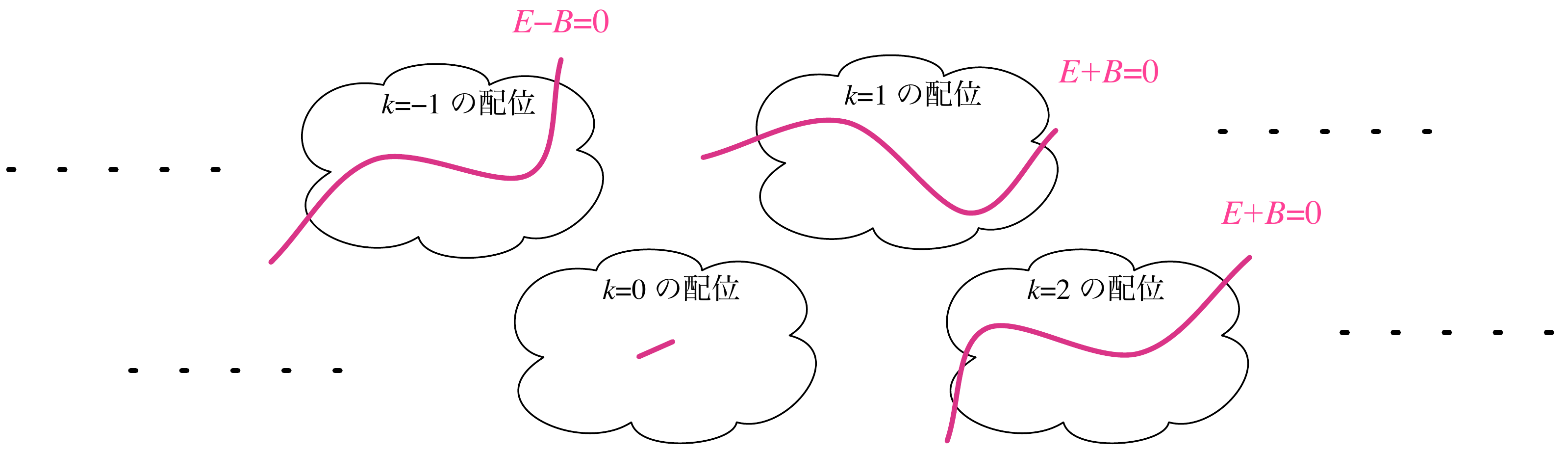}
\]
\caption{無限次元の配位空間は、インスタントン数 $k$ でラベルされる部分空間にわかれる。そのそれぞれの中で、(反)自己双対解は作用を最小化する。\label{components}}
\end{figure}

さて、
\begin{align}
-\tr F_{\mu\nu}F_{\mu\nu} &=  -2\tr (\vec E^2+\vec B^2) \\
&=-2\tr( \vec E\pm \vec B)^2 \pm 4 \tr \vec E\cdot\vec B \\
&\ge \pm 4 \tr \vec E\cdot\vec B  = \pm\tr F_{\mu\nu}\tilde F_{\mu\nu}
\end{align} ですから、
\begin{equation}
\int d^4 x \tr F_{\mu\nu} F_{\mu\nu} \ge \left|\int d^4 x \tr F_{\mu\nu} \tilde F_{\mu\nu}\right|  = 16\pi^2 |k|\hbox{、}
\end{equation}  この等号を満たすには $k$ が正ならば \begin{equation}
\vec B+\vec E=0, \quad\hbox{もしくは}\quad F_{\mu\nu} + \tilde F_{\mu\nu}=0 \label{SD}
\end{equation} であればよいことがわかります(再度図\ref{components}を参照下さい)。これを反自己双対方程式といいます。
$k$ が負の場合は $F_{\mu\nu}=\tilde F_{\mu\nu}$ という自己双対方程式を考えれば良いですが、本質的に同じですので、今後は $k$ は非負としましょう。

さて、反自己双対方程式を調べる前に、$k$ が整数になることを確かめましょう。まず、$\tr F_{\mu\nu} \tilde F_{\mu\nu}$ が全微分であることに注意します: \begin{equation}
\tr F_{\mu\nu} \tilde F_{\mu\nu} = \partial_{\mu} \epsilon_{\mu\nu\rho\sigma} \tr( A_\nu F_{\rho\sigma} -\frac13 A_\nu A_\rho A_\sigma)\hbox{。} 
\end{equation} よって、\begin{equation}
-\int d^4x  \tr F_{\mu\nu} \tilde F_{\mu\nu} 
= \int_{S^3} dn_\mu \epsilon_{\mu\nu\rho\sigma} \tr ( -A_\nu F_{\rho\sigma} + \frac13 A_\nu A_\rho A_\sigma)
\end{equation} です。無限遠で $F_{\mu\nu}=0$ としましたから、ゲージ変換 \eqref{YMgauge} を $F_{\mu\nu}=0$、 $A_{\mu}=0$ から逆に使って、\begin{equation}
= \frac13\int_{S^3} dn_\mu \epsilon_{\mu\nu\rho\sigma} (g^{-1}\partial_\nu g)(g^{-1}\partial_\rho g)(g^{-1}\partial_\sigma g)\hbox{、}
\end{equation} ただし、$g$ は無限遠の $S^3$ から群 $\SU(N)$ への写像です。もっとも簡単な $N=2$ のときを考えますと、$\SU(2)\sim S^3$ ですから、$g$ は \begin{equation}
g: S^3\to S^3
\end{equation} と思うことができ、上記積分はこの写像が何回巻き付いているかを測ったものになります。比例係数を計算すると、まきつき数を $k$ として\begin{equation}
= 16\pi^2 k\hbox{。}
\end{equation}

反自己双対方程式の良いところは、まず、これを満たせば自動的に Yang-Mills 方程式を満たすことがあります: \begin{equation}
\partial_\mu F_{\mu\nu}+ [A_\mu, F_{\mu\nu}] = -\partial_\mu \tilde F_{\mu\nu} - [A_\mu,\tilde F_{\mu\nu}] 
\end{equation}ですが、右辺に $F_{\mu\nu}$ の定義を代入すると自動的にゼロになることがわかります。
また、Yang-Mills 方程式は二階の微分方程式ですが、反自己双対方程式は一階です。さらに、非線形項も三次でなくて二次でおさまります。

\subsubsection{1-インスタントン解}
では、インスタントン数が $k$ の反自己双対解はどんな形をしているのでしょうか? まず一番簡単な $k=1$ の場合を見ましょう。これは一般に \begin{equation}
A_\mu(x) = \frac{H_{\mu\nu} (x-x_0)_\nu}{|x-x_0|^2 +\rho^2}
\end{equation} という形をしていると知られています。ここで $x_0$ はインスタントンの中心、$\rho$ はインスタントンの大きさを決めます。
$H_{\mu\nu}$ はどう取ればいいでしょうか? 
答えを反自己双対にしたいので、$H_{\mu\nu}=-H_{\nu\mu}$、$H_{\mu\nu}=-\tilde H_{\mu\nu}$ という
$N\times N$ 行列をとることにします。この形を反自己双対方程式に代入すると、$H_{01}$、 $H_{02}$、 $H_{03}$ が \begin{equation}
H_{01}=[H_{02},H_{03}]\hbox{、} \quad
H_{02}=[H_{03},H_{01}]\hbox{、}\quad
H_{03}=[H_{01},H_{02}]
\end{equation} と、$\SO(3)$ の交換関係をみたすべきことがわかります。さらに、インスタントン数を計算すると \begin{equation}
k=2\tr H_{03}{}^2
\end{equation} となることもわかります。ですから、一番簡単な解は、\begin{equation}
H_{0i}=\frac12 i\sigma_i \oplus O_{N-2}
\end{equation}とすることです。但し $\sigma_{1,2,3}$ は通常のパウリ行列 \begin{equation}
\sigma_1=\begin{pmatrix}
0&1 \\
1&0
\end{pmatrix}, \quad 
\sigma_2=\begin{pmatrix}
0&i \\
i&0
\end{pmatrix}, \quad 
\sigma_3=\begin{pmatrix}
1&0 \\
0&-1
\end{pmatrix}
\end{equation}で、これは $2\times 2$ 行列ですから、$O_{N-2}$ は$(N-2)\times (N-2)$ 行列で全てゼロなものとして、それを付け足して $N\times N$ 行列にすることにします。$\oplus$ は行列をブロック対角にならべる操作です。

ひとつ解ができると、勝手な $\SU(N)$ 行列 $g$ を取って \begin{equation}
H_{0i}=g^{-1}(\frac12 i  \sigma_i \oplus O_{N-2}) g\label{g}
\end{equation}としても当然解になります。$g$ は $ (h \, \mathrm{Id}_{2\times 2}) \oplus g_{(N-2)\times (N-2)} $ という形の行列だと空回りしますので、これで \begin{equation}
N^2-1-(N-2)^2= 4N-5
\end{equation} 自由度だけ解が得られたことになります。$x_0$ にある四つの自由度および $\rho$ にある一つの自由度を足すと、合計 $4N$ 自由度あることがわかりました。
これらの自由度のことをインスタントンのモジュライと呼びます。

特に $N=2$ の場合は、$4\cdot 2-5=3$ 自由度は $\SU(2)$ 行列 $g$ でまわす自由度そのものです。この自由度は $\SU(2)\simeq S^3$ だけあります。ただし、\begin{equation}
g=\begin{pmatrix}
-1 & 0\\
0& -1
\end{pmatrix}
\end{equation} の場合は交換してしまって $H_{\mu\nu}$ に影響がないので、実際に意味があるのは $\SU(2) / \{\pm 1\} \simeq S^3/\bZ_2$ だけです。 結論として、$\SU(2)$ の1-インスタントン解のモジュライの空間 $\cM_{2,1}$ は \begin{equation}
\cM_{2,1}\simeq \bR^4 \times \bR_+ \times S^3/\bZ_2 \simeq \bR^4 \times \bR^4/\bZ_2
\end{equation} であたえられることがわかります。但し、右辺では大きさのパラメタ $\rho$ とゲージの向きのパラメタ $S^3/\bZ_2$ をくみあわせて $\bR^4/\bZ_2$ としました。

\subsubsection{多重インスタントン解}
さて、前節では 1-インスタントン解には中心の位置に $4$ パラメタ、サイズに $1$ パラメタ、ゲージの向きに $4N-5$ パラメタあることを学びました。具体的に作用密度 $\tr F_{\mu\nu}^2$ を図示すると図~\ref{instanton} のようになります。これからもわかるように、中心から離れるとゲージ場の強さ $F_{\mu\nu}$ は十分に小さくなります。

\begin{figure}
\[
\includegraphics[width=.3\textwidth]{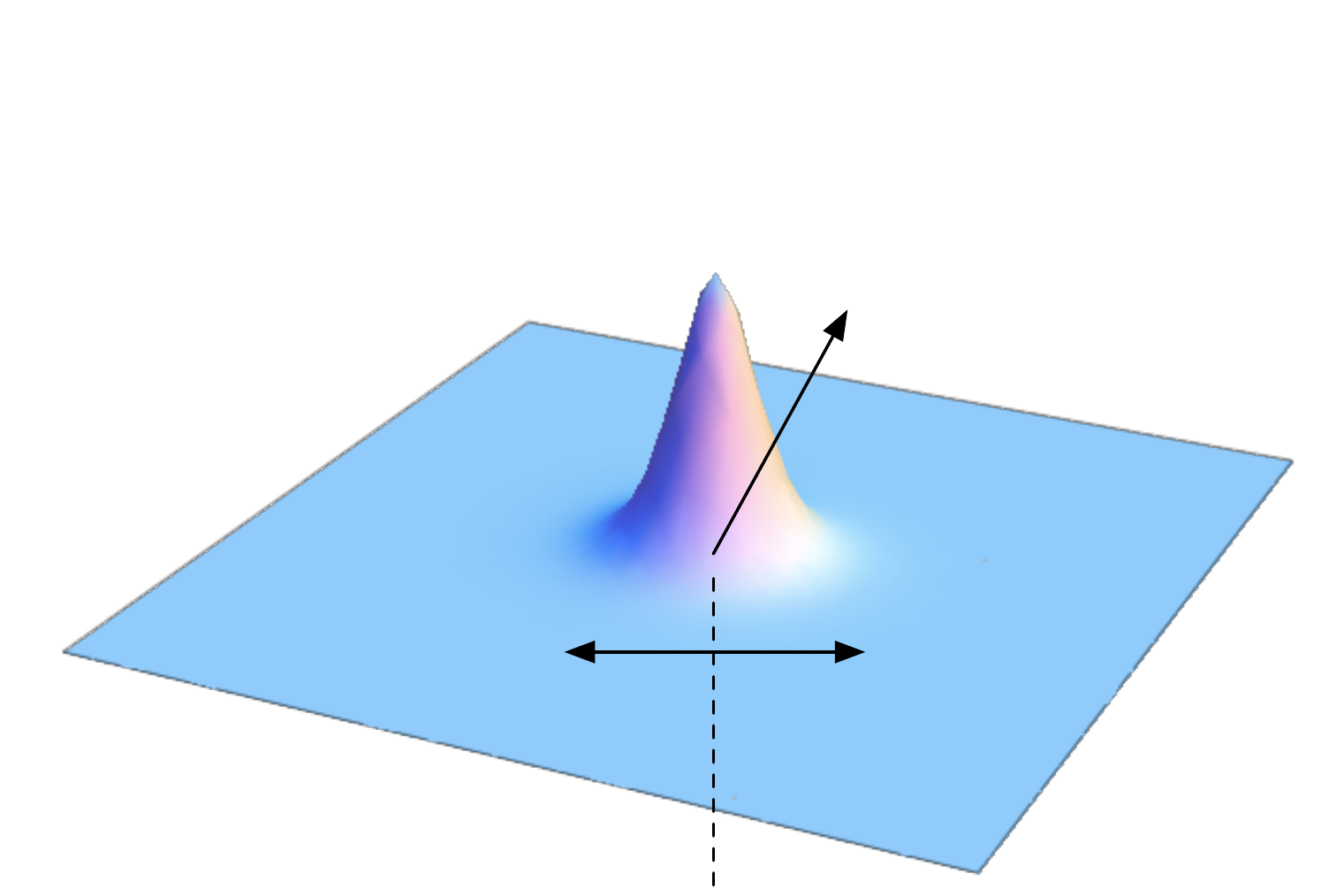}
\]
\caption{1-インスタントン解の作用密度\label{instanton}}
\end{figure}

\begin{figure}
\[
\begin{array}{cc}
\raisebox{0cm}{\includegraphics[width=.3\textwidth]{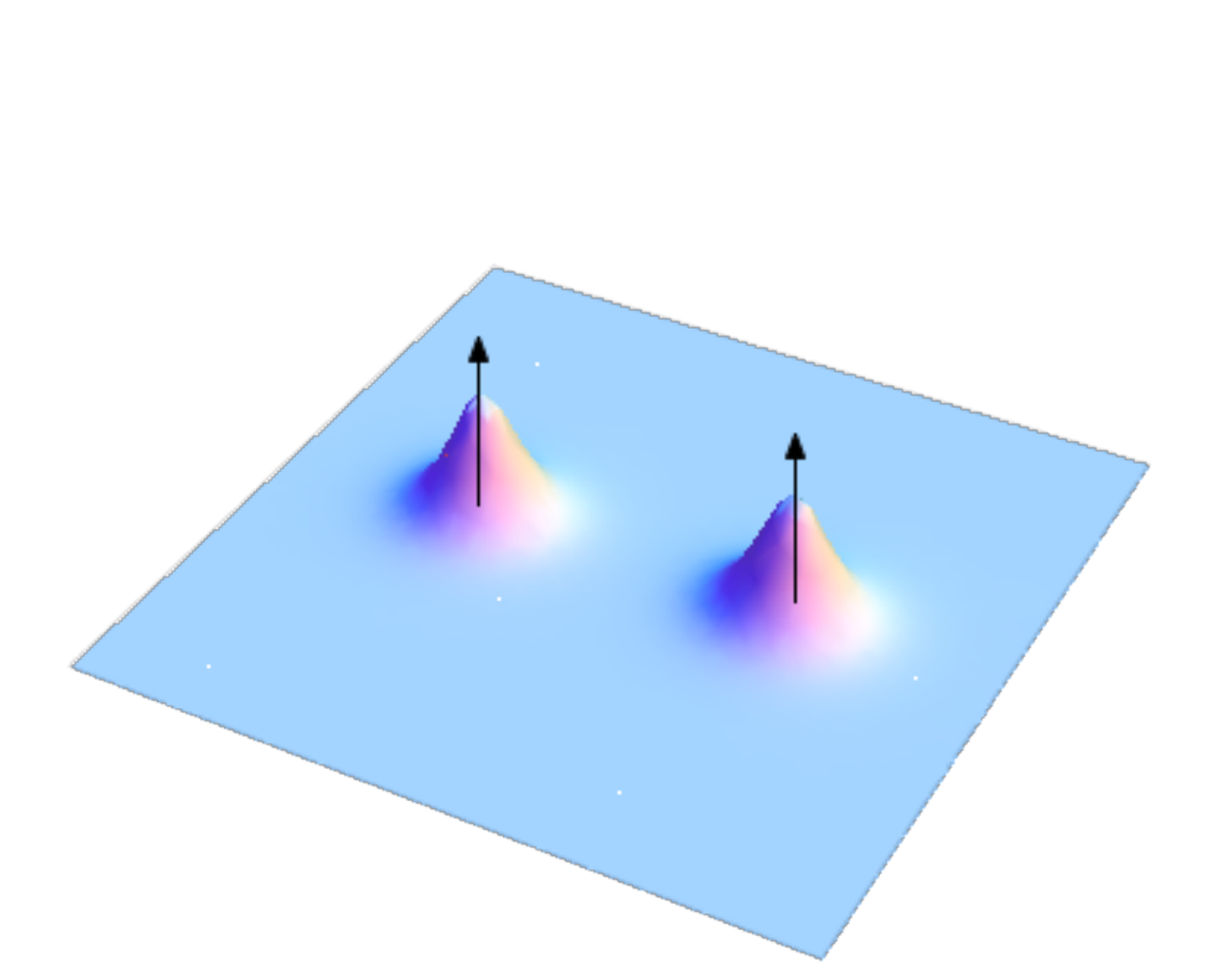}}&
\raisebox{0cm}{\includegraphics[width=.3\textwidth]{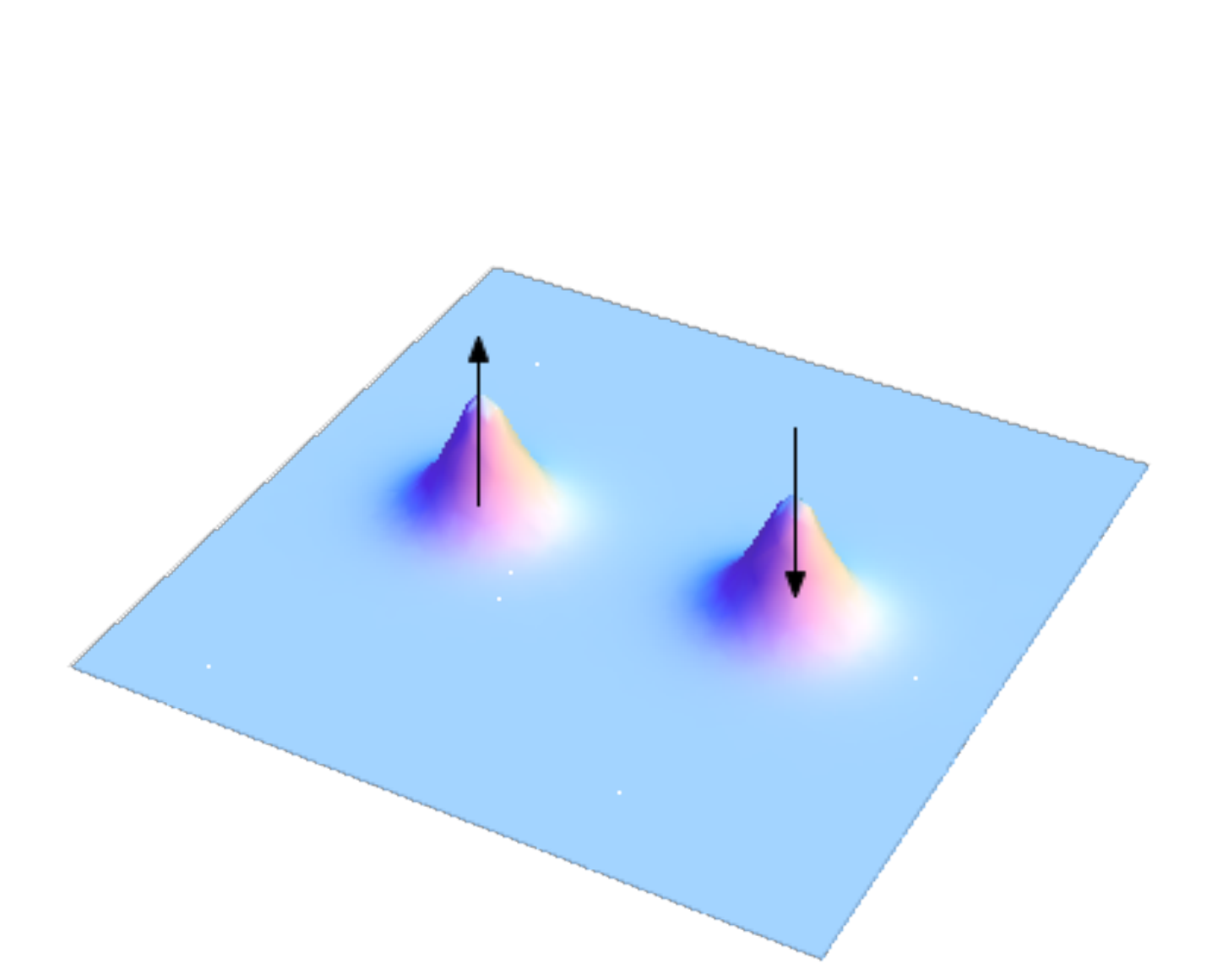}}\\
\downarrow & \downarrow \\
\raisebox{0cm}{\includegraphics[width=.3\textwidth]{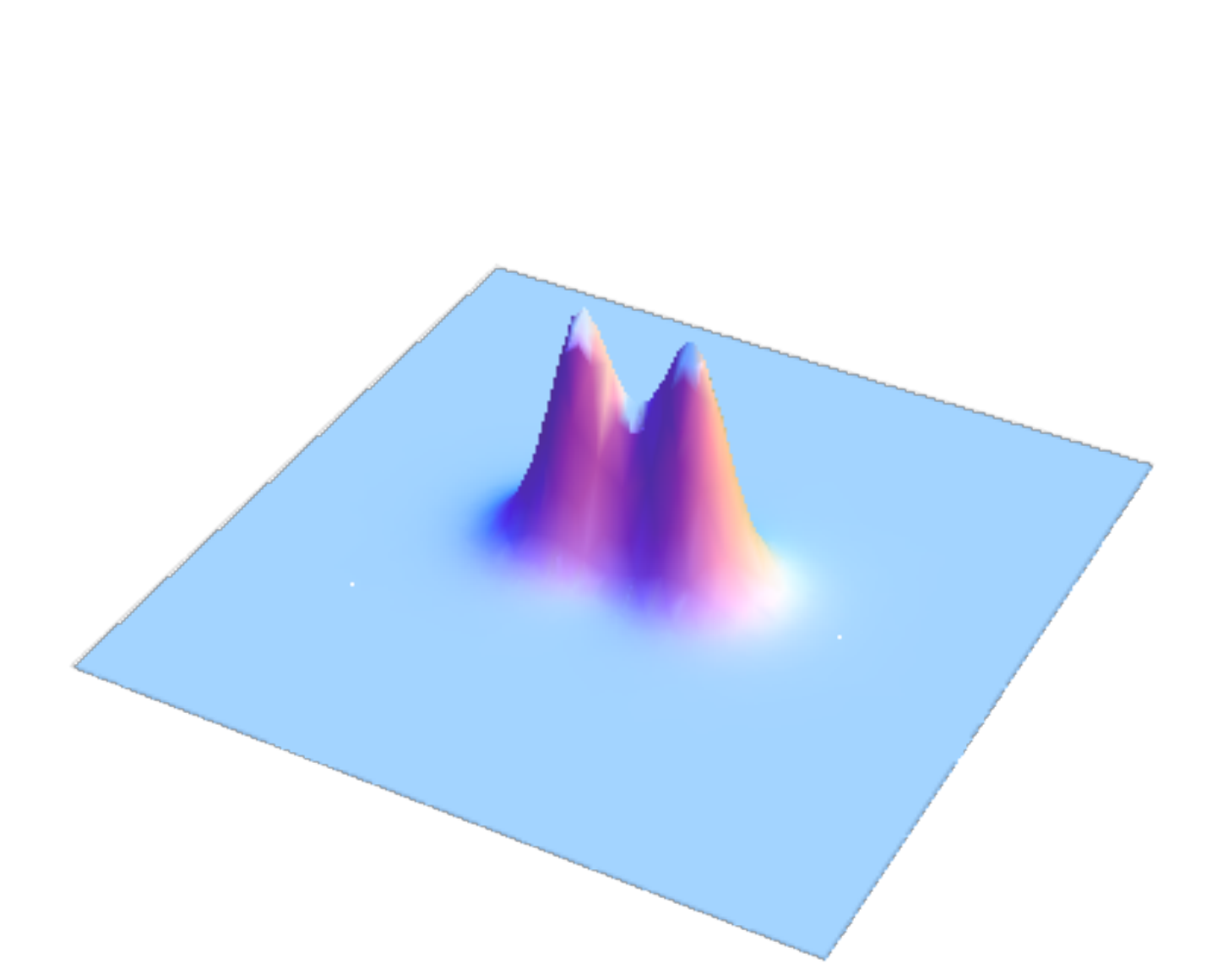}}&
\raisebox{0cm}{\includegraphics[width=.3\textwidth]{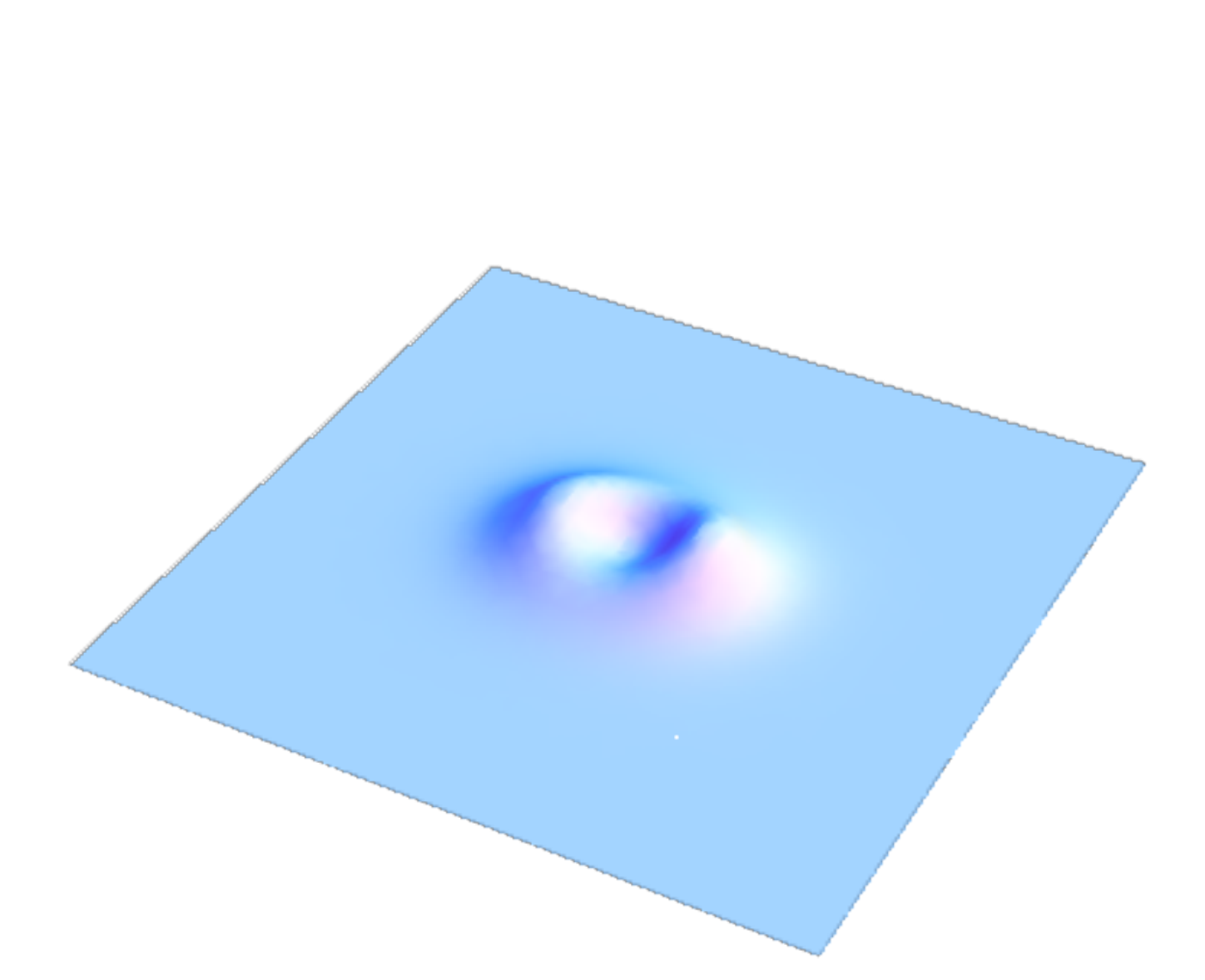}}
\end{array}
\]
\caption{2-インスタントン解の作用密度。二つのインスタントンを近づける際、ゲージ場の向きがどうなっているかによって結果は著しく異なる。\label{2instanton}}
\end{figure}

ですから、1-インスタントン解をふたつ、片方 $A'_\mu$ は中心が $x_0'$ サイズが $\rho'$、 もうひとつ $A_\mu''$は中心が $x_0''$ サイズが $\rho''$ であるように取ると、
$|x_0'-x_0''| \gg \rho'+\rho''$ であれば、両者が同時に大きな値にはなりませんから、$A_\mu' + A_\mu''$ もほとんど反自己双対方程式の解になります。
勿論 $A_\mu' A_\mu''$ からくる補正があるので、それを修正してやらないといけません: \begin{equation}
A_\mu = A_\mu' + A_\mu'' + \text{小さな補正}\hbox{。}
\end{equation} この領域では、2-インスタントン解には $8N$ パラメタがあることがわかります。インスタントンが二つ近づいた場合は、このような安直な解析ではいけませんが、その場合でも方程式をきちんと解けることが知られています(図\ref{2instanton})。

一般にインスタントン数が $k$ の場合は、同様にして全てのインスタントンの中心が離れていると、重ね合わせることによって $4Nk$個モジュライをもった解がつくれます。
いいかえると、このモジュライ空間を $\cM_{N,k}$ と書くと、モジュライ空間の外の方ではおおよそ
\begin{equation}
\cM_{N,k} \sim (\cM_{N,1})^k / S_k
\end{equation} となっている、すなわち 1-インスタントンのモジュライのコピーが $k$ 個あってそれを置換群 $S_k$ で同一視したものになっていますが、中心部はもっと複雑になっています。

$\cM_{N,k}$ を具体的に書き下す方法も  Atiyah-Drinfeld-Hitchin-Manin \cite{Atiyah:1978ri} によって示されています。それをきちんと説明する時間は到底ありませんが、雰囲気だけは説明したいと思います(この節の詳細はレビュー \cite{Dorey:2002ik} や講義録 \cite{Ito} 等を参照のこと。)。
まず、$k=1$ の場合に戻りますと、解の微妙な部分は $H_{0i}$ で与えられていました。$H_{0i}$ は、$X=H_{01}+iH_{02}$ で から復元できますが、条件 $X^2=0$ と $\tr |X|^2=1$ が必要です。ここで、 $\rho$ の自由度を $X$ に含めてしまえば、$\tr |X|^2 $に関する条件は落とすことができます。そこで、$X^2=0$ にだけ注目しましょう。すると、$X$ の階数は $1$ なので、\begin{equation}
X^i_j=B^iA_j 
\end{equation}  と書く事ができます。但し $i,j=1,\ldots,N$。 $X^2=0$ を満足させるために 
\begin{equation}
A_i B^i=0 \label{foo}
\end{equation} を要求して、さらに勝手な $c\in \bC$ に対して $(A_i,B_i) \to (cA_i,c^{-1} B_i)$ としても $X$ が変わらないので、\begin{equation}
A_i \overline{A}^i -B^j \overline{B}_j =0\label{bar}
\end{equation}としてせめて$c\in \bR$ の自由度は消すことにしましょう。おしまいに、中心の自由度 $x_0\in \bR^4$ を $(z,w)\in \bC^2$ と書くとすると、結局 1-インスタントン解は $(z, w, A_i,B^i)$ で \eqref{foo}、 \eqref{bar} を満たし、さらに 
\begin{equation}
(A_i,B^i)\to (e^{i\theta} A_i,e^{-i\theta}B^i) \label{baz}
\end{equation}
という作用に対して同一視をしたもの、と書く事ができます。
Atiyah-Drinfeld-Hitchin-Manin は、これの自然な拡張が $k$-インスタントン解のモジュライを記述する事を見いだしました。
答えだけ書きますと、$1,\ldots,k$ を走る添字 $a,b$ を用意し、上記の $z,w,A,B$ に添字を追加して $z^a_b$, $w^a_b$, $A_i^a$, $B^j_a$  とします。そうして、\eqref{foo} の拡張として \begin{equation}
A_i^a B^i_b +[z,w]^a_b =0\hbox{、} \label{FOO}
\end{equation} \eqref{bar} の拡張として \begin{equation}
A_i^a \overline{A}^i_b - B^i_b \overline{B}^a_i + [z,z^\dagger]^a_b + [w,w^\dagger]^a_b=0\label{BAR}
\end{equation} を課し、 \eqref{baz} の拡張として、\begin{equation}
(A_i^a,B^i_b,z,w)\to (g^a_b A_i^b,B^i_a (g^{-1})^a_b,  gzg^{-1}, gwg^{-1}) \label{BAZ}
\end{equation} という $k\times k$ ユニタリ行列 $g^a_b$ の作用で同一視することにせよ、というのが彼らの見つけた表示です。

モジュライの数を勘定しましょう。$A$\hbox{、} $B$ には合計 $4Nk$ 自由度があり、$z$\hbox{、} $w$  には $4k^2$ 個自由度があります。 \eqref{FOO} で $2k^2$ 個条件を課し、\eqref{BAR} で $k^2$ 個、さらに \eqref{BAZ} で $k^2$ 個自由度を取り除くので、結局 $4Nk$ 個自由度があることになります。
1-インスタントン解を $k$ 個とってきた場合も、対応する $(A_i,B^i,z,w)$ を $k$ 組とってくれば、\eqref{FOO}\hbox{、} \eqref{BAZ} をおおよそ解く行列を作るのはブロック対角に並べればいいですが、$[z,w]$ の交換子のあたりからインスタントン間の相互作用が出てくるわけです。

\subsection{インスタントンの統計力学}
さて、インスタントン解について多少学んだところで、経路積分の評価にもどりましょう。作用はインスタントン数が $k$ の配位の中では反自己双対な配位で最小になり、経路積分にもっとも寄与するのですから、一般の配位 $A_\mu$ を \begin{equation}
A_\mu = A_\mu^\text{ASD} + \delta A_\mu 
\end{equation}のように反自己双対部分 $A_\mu^\text{ASD}$ とそこからのずれ $\delta A_\mu$ に分解することを考えます(図\ref{decomp})。

\begin{figure}
\[
\includegraphics[width=.3\textwidth]{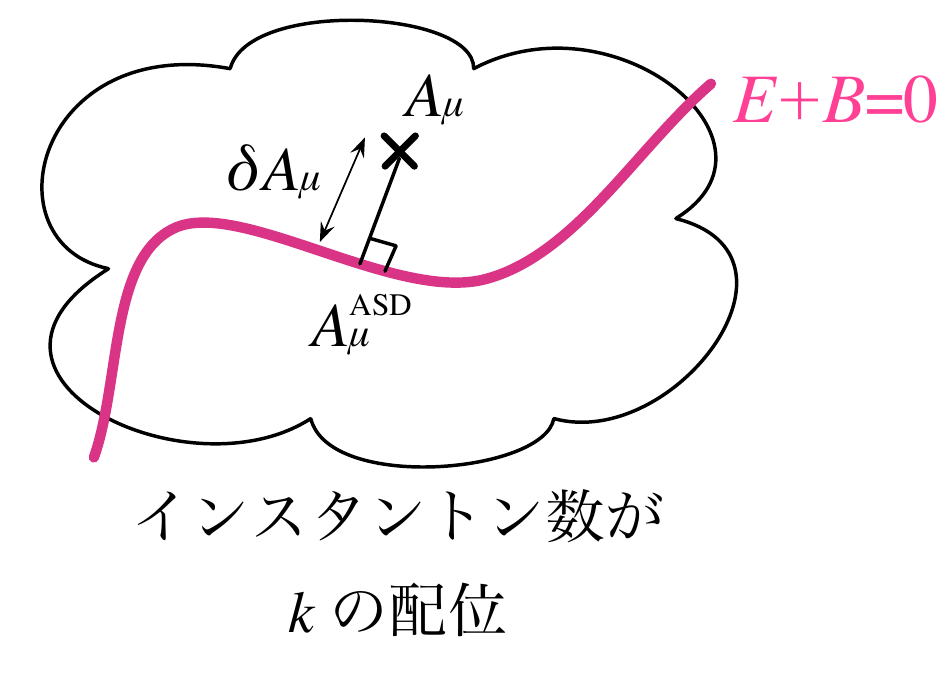}
\]
\caption{一般の配位 $A_\mu$を反自己双対部分 $A_\mu^\text{ASD}$  とそこからのずれ $\delta A_\mu$ に分解する。\label{decomp}}
\end{figure}

$\delta A_\mu$ が小さければ、\begin{equation}
S = \frac{8\pi^2 k}{ g^2} + \int d^4 x \text{($\delta A_\mu$の二次式)} + \text{(高次項)}
\end{equation}となり、経路積分の積分変数も分解すると \begin{equation}
Z=\int [DA_\mu] e^{-S} =\sum_k \int [DA_\mu^\text{ASD}] \int [\delta A_\mu] e^{-\frac{8\pi^2  k}{g^2} + \cdots } 
\end{equation}となります。この展開を利用して量子 Yang-Mills 理論を調べようとはじめにがんばったのが 't Hooft の論文\cite{'tHooft:1976fv}ですが、揺らぎ $\delta A_\mu$ の処理は非常に大変なので、今回はそれをすっかり省略して、以下のトイモデルを考えましょう: \begin{equation}
Z_\text{toy}^\text{instanton}=\sum_k  q^k \int [DA_\mu^\text{SD}] = \sum_k q^k \int_{\cM_{N,k}} d\vol\hbox{、}
\end{equation} 但し $d\vol$ $\cM_{N,k}$ の上の自然な体積形式です。
$q=\exp(-{8\pi/ g^2})$ はインスタントン数の化学ポテンシャル、すなわちをインスタントンを一つ系に導入する際のコストだと思う事ができます。
あとは自然な体積形式以外は何も積分していませんから、どのパラメタの反自己双対配位も同じ確率で起こりうるという状況を考えています。

勿論このままでは $Z$ は時空の積分のために発散します。すなわち、$d^{4Nk}s$ の中には、インスタントンの中心の位置 $\bR^4$ に関する積分があるので、その分の発散があります。統計力学を学びますと、通常この問題は時空を一辺 $L$ の大きな超立方体の箱に入れて計算して、$\log Z \sim L^4$ となるその比例定数を取りだすことで処理しますが、インスタントンは箱にいれると更に解析が難しくなるという性質がありますので、ちょっと別のことをしてみます。

そのために、もっと簡単な点粒子の模型を考えましょう、単に粒子が $(x,y)\in \bR^2$ のどの箇所にも同じ確率で存在しうるとします。すると、勿論分配関数は \begin{equation}
Z=\int_{-\infty}^\infty\int_{-\infty}^\infty dxdy =\infty
\end{equation}となって発散します。そのかわりに、ガウス型の因子を手でいれて収束させましょう: \begin{equation}
Z_{\epsilon}= \int_{-\infty}^\infty \int_{-\infty}^\infty e^{-\pi \epsilon (x^2+y^2)}dxdy = \frac1{\epsilon}\hbox{。}
\end{equation} $\epsilon\to 0$ とすると収束因子を取り払うことになって、それにともなって $Z$ も発散します。ですから $1/\epsilon$ は収束因子を入れた際の実効的な $\bR^2$ の面積と思うことが出来ます。

これはどうみても手でむりやり収束させたような感じですが、もうすこし意味付けをすることができます。$(x,y)$ は空間の場所としましたが、ポアソン括弧を入れて力学系の相空間の座標と思うことにします: \begin{equation}
\{x,y\}_\text{P.B}=1\hbox{。}
\end{equation} そうすると、$J=(x^2+y^2)/2$ と書くと、\begin{equation}
\{H,x\}_\text{P.B.}=y\hbox{、}\quad
\{H,y\}_\text{P.B.}=-x
\end{equation}ですから、$J$ は $(x,y)$ 平面の回転のハミルトニアンですね(図\ref{harmonic})。ですから、上記の $Z_\epsilon$ は、 \begin{equation}
Z_\epsilon = \int_{-\infty}^\infty \int_{-\infty}^\infty e^{-2\pi \epsilon J}dxdy
\end{equation} と、回転のハミルトニアン $J$ が大きくなると $\epsilon$ の重みで損をするように積分しなさい、ということだと思えます。今回はこの方法で系を箱に入れましょう。

\begin{figure}
\[
\includegraphics[width=.3\textwidth]{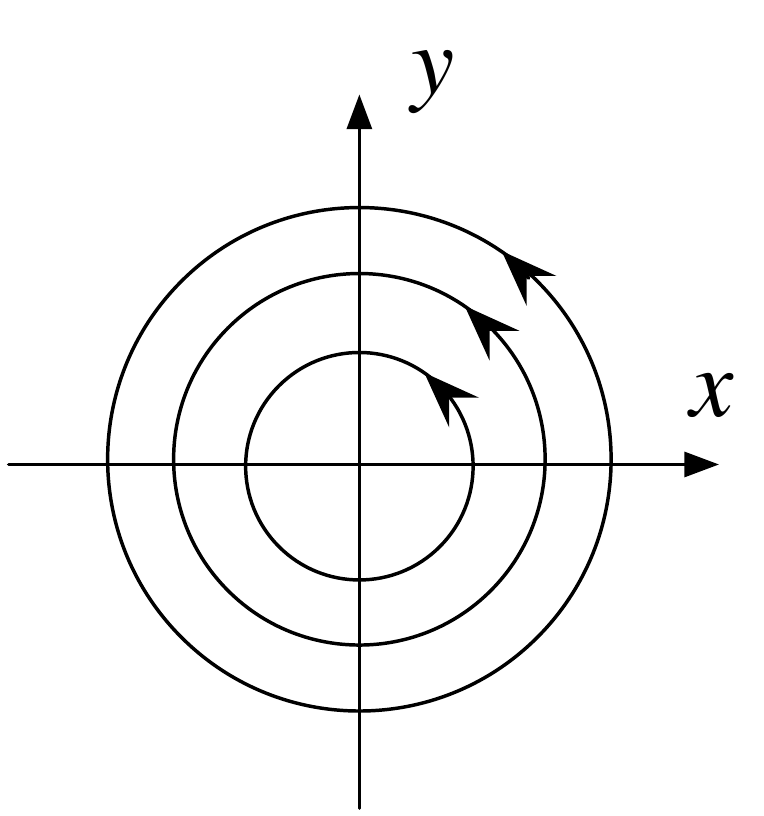}
\]
\caption{$J=x^2+y^2$ は相空間の回転を導く。\label{harmonic}}
\end{figure}

インスタントンに話を戻しますと、今は時空は四次元で $x_1,x_2,x_3,x_4$ とありますから、ポアソン括弧を \begin{equation}
\{x_1,x_2\}_\text{P.B.}=1\hbox{、}\qquad
\{x_3,x_4\}_\text{P.B.}=1
\end{equation} その他はゼロ、として位相空間と思うことにしましょう。このように $\bR^4$ に位相空間の構造をいれると、自然にインスタントンのモジュライ空間 $\cM_{N,k}$ にも相空間の構造が入ることが知られています。
$\cM_{N,k}$ の一点をとりましょう。すると、ひとつ $A_\mu(x)$ というインスタントン配位が定まりますから、 $\bR^4$ の回転をすると、$A_\mu'(x)$ という別個のインスタントン配位が定まり、$\cM_{N,k}$ の別の一点になります。ですから、$\bR^4$ の回転は自然に $\cM_{N,k}$ に働きます。
そこで、$\bR^4$ の $(x_1,x_2)$ 平面を回す回転が $\cM_{N,k}$ に引き起こす作用を考え、これのハミルトニアンを $J_1$, $\bR^4$ の $(x_3,x_4)$ を回す回転に対応するハミルトニアンを $J_2$ と呼ぶことにしましょう。それぞれに対する重みを $\epsilon_{1,2}$ と呼ぶと、\begin{equation}
e^{-2\pi (\epsilon_1 J_1 + \epsilon_2 J_2)}
\end{equation} という重みをつけることにします。

また、大域ゲージ変換も $\cM_{N,k}$ の変換になります。簡単のため $\SU(2)$ を考えると、$\sigma_3$ で生成されるゲージ変換に対応するハミルトニアン $K$ を考え、重み $a$ をつけましょう。すると、指数関数の肩に $aK$ という因子をいれることになります。

一般の $\SU(N)$ では、対角行列のゲージ変換 $\diag(a_1,\ldots,a_N)$ に対して、ハミルトニアン $K_1,\ldots, K_N$ を考え、指数関数の肩に $\sum_ia_i K_i$ という因子をいれることにします。

というわけで、我々が考えたいインスタントンの統計力学模型は次のようなものです: \begin{equation}
Z^\text{instanton}_{\epsilon_1,\epsilon_2;a_i}  = \sum_k q^k Z_{N,k}\hbox{、} \quad \text{但し}\quad Z_{N,k}=\int_{\cM_{N,k}} e^{-2\pi(\epsilon_1 J_1 + \epsilon_2 J_2 + \sum_i a_i K_i)} d\vol\hbox{。}
\end{equation}

随分ややこしいことを言いましたが、結局は、インスタントンを沢山いれると $q$ だけ損をする、 $(x_1,x_2)$ 平面内の角運動量が大きいと $\epsilon_1 J_1$ だけ損をする、
$(x_3,x_4)$ 平面内の角運動量が大きいと $\epsilon_2 J_2$ だけ損をする、
ゲージ変換性が大きいと パラメタ $a_i$ に応じて $\sum_i a_i K_i$  だけ損をする、という、インスタントンが沢山あったばあいに統計力学的に扱おうとする際に考えることの出来るもっとも簡単な模型になっています。$J_1$、 $J_2$、 $K_i$ も具体形は結局は単に調和ポテンシャルになるだけです。

それを実際に $\SU(2)$ の 1-インスタントンの場合に確かめてみましょう。モジュライ空間は先に $\bR^4 \times \bR^4/\bZ_2$ だといいました。ひとつめの $\bR^4$ はインスタントンの中心 $(x_1,x_2,x_3,x_4)$ をパラメタしていて、ふたつめの $\bR^4$ はゲージの向きをあらわす行列 $H_{0i}$ と インスタントンのサイズ $\rho$ をあわせたものでした。
ふたつめの $\bR^4/\bZ_2$ を複素数ふたつ $(z,w)$ を $(z,w)\to -(z,w)$で同一視したものと思うことにしましょう。

ゲージ回転は単に $(z,w)$ が$\SU(2)$ の二次元表現ですから \begin{equation}
(z,w)\mapsto (e^{ia}z,e^{-ia} w)
\end{equation} と働きます。一方、
$x_1,x_2$ 平面の $\epsilon_1$ 回転および $x_3,x_4$ 平面の $\epsilon_2$ 回転は \begin{equation}
(x_1+ ix_2,x_3+ix_4,z,w) \mapsto (e^{i\epsilon_1}(x_1+ix_2),  e^{i\epsilon_2}(x_3+ix_4), e^{i(\epsilon_1+\epsilon_2)/2}z,e^{i(\epsilon_1+\epsilon_2)/2}w)
\end{equation} と作用します。ここで、$(z,w)$ への作用は、時空の回転が $H_{\mu\nu}$ を混ぜることから、式 \eqref{g} にあらわれる $g$ を変換する必要があり、$g$ が $(z,w)\in\bC^2$ の角度部分であったことから生じます。

以上から、対応するハミルトニアンは \begin{multline}
\epsilon_1 J_1 + \epsilon_2 J_2 + a K = \frac{\epsilon_1}2 (x_1^2+x_2^2)+\frac{\epsilon_2}2 (x_3^2+x_4^2)\\
+ \frac{(\epsilon_1+\epsilon_2)/2+a}2 |z|^2 +  \frac{(\epsilon_1+\epsilon_2)/2-a}2 |w|^2 
\end{multline} ですね。
すると \begin{equation}
Z_{2,1}
= \frac12 \frac{1}{\epsilon_1}\frac{1}{\epsilon_2} \frac{1}{(\epsilon_1+\epsilon_2)/2 +a} \frac{1}{(\epsilon_1+\epsilon_2)/2 -a}\label{M21}
\end{equation} となります。先頭の $1/2$  は $\bR^4/\bZ_2$ の $/\bZ_2$ から来ます。

すこしだけこの計算結果の解釈をしておきましょう。$1/(\epsilon_1\epsilon_2)$ は時空の箱の大きさですから、箱を大きくするために $\epsilon_{1,2}\to 0$ とします。するとガウス積分 \eqref{M21} の収束性が悪くなりますから、$a$ は純虚として計算が問題ないようにします。
すると、時空の箱が大きい極限では、インスタントンが一個あることによる効果は単位時空体積あたり \begin{equation}
\epsilon_1\epsilon_2\log Z_{2,1}\sim \frac12 \frac{1}{(\epsilon_1+\epsilon_2)/2 +a} \frac{1}{(\epsilon_1+\epsilon_2)/2 -a}
\sim -\frac12\frac1{a^2}
\end{equation} となるわけです、ただし $|a|\gg |\epsilon_{1,2}|$ としました。

\subsection{積分の局所化}
さて、$\SU(2)$ の場合の $Z^\text{instanton}$ を計算するには、この調子で $\cM_{2,2}$, $\cM_{2,3}$, \ldots 上の積分を計算できればいいのですが、それを直接行うのはなかなか大変です。そこで、Duistermaat-Heckman の局所化公式というものをつかいます。(この節の内容の詳細は、教科書\cite{GS}を参照のこと。)

この公式は、丁度我々が計算したいような、ハミルトニアンを指数関数の肩にのせたものの積分をたちどころに出来てしまうものです。一番簡単な例として、二次元球面を考えます。緯度を $-\frac\pi2<\theta<\frac\pi 2$、 経度を $0<\psi<2\pi$ として、面要素を $\cos\theta d\theta d\psi$ とします。経度方向回転のハミルトニアンは $H=\sin\theta $です。そこで \begin{equation}
Z=\iint e^{-2\pi\epsilon H} \cos\theta d\theta d\psi 
\end{equation}を考えましょう。これは直接計算できて、 \begin{equation}
=\frac{e^{-H(\theta=\pi/2)}}{\epsilon}-\frac{e^{-H(\theta=-\pi/2)}}{\epsilon}
\end{equation} となりますが、これを次のように書きます\footnote{一変数関数の定積分の公式が固定点定理の一種であるというのは講演 \cite{NakajimaSougou} から学びました。} : \begin{equation}
=\sum_{p=\text{北極,南極}} \frac{e^{-H(p)}}{\text{($p$ での回転角)}}\hbox{。}
\end{equation}

Duistermaat-Heckman の公式は、これが一般になりたつ、というもので、$M$ が $2n$次元の滑らかな相空間で、$H$ がハミルトニアンとしてその上に流れを引き起こし、流れの固定点 $p$ が孤立しているとすると、\begin{equation}
Z=\int e^{-2\pi\epsilon H} d\mathrm{vol} = \sum_p \frac{e^{-H(p)}}{\prod_{i=1}^{n} \theta_{i,p}}\label{DH}
\end{equation} が成り立つ、というものです。ただし、各固定点 $p$ のまわりでは、流れは点 $p$ まわりの回転になりますから、$\bR^{2n}$ を $n$ 個の $\bR^2$ にわけてそれぞれの面が角度 $\theta_{i,p}$  ($i=1,\ldots,n$)で回転している、というようにしました。

もうひとつ実例を見ましょう。 $M=\bR^4$ として、$(x_1,x_2)$ を角速度 $\epsilon_1$ で、$(x_3,x_4)$ を角速度 $\epsilon_2$ で回すことを考えると、ハミルトニアンは \begin{equation}
J=\frac{\epsilon_1}2(x_1^2+x_2^2)+\frac{\epsilon_2}2(x_3^2+x_4^2)
\end{equation}です。公式を適用するには、固定点を探さねばなりませんが、勿論それは原点だけで、そこでの回転角は $\epsilon_1$ と $\epsilon_2$ ですね。よって\begin{equation}
\int_{-\infty}^\infty\int_{-\infty}^\infty\int_{-\infty}^\infty\int_{-\infty}^\infty 
e^{-2\pi J }d^4 x = \frac{1}{\epsilon_1\epsilon_2}
\end{equation}となるはずですが、これは単にガウス積分です。これは先ほどやった積分の $\bR^4$ 部分です。

この計算を $\bR^4/\bZ_2$ に適用するにはどうすればよいでしょうか? ひとつ問題は、定理は滑らかな多様体に対してのみ成り立つのですが、$\bR^4/\bZ_2$ は先端がとがっています。
とがりを解消する為、ブローアップという操作をします。まず、$(z,w)\in\bC^2\simeq \bR^4$  を座標として、$(z,w)\mapsto -(z,w)$ という $\bZ_2$ 作用で割っていますから、$u=z^2$, $v=w^2$, $t=zw$ が $\bZ_2$ 作用で不変です。これらは $uv=t^2$ を満たします。$u=v=t=0$ のあたりが非常に尖っています。

\begin{figure}\[
\begin{array}{c@{\qquad}c}
\includegraphics[width=.3\textwidth]{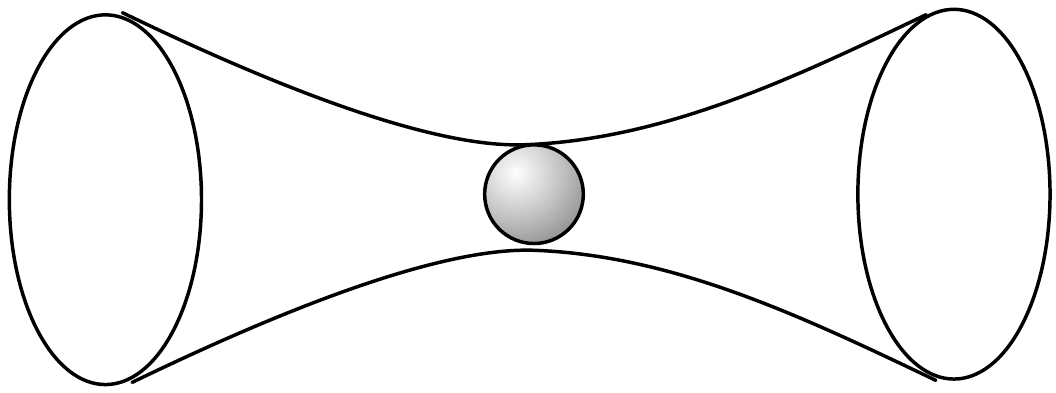}&
\includegraphics[width=.3\textwidth]{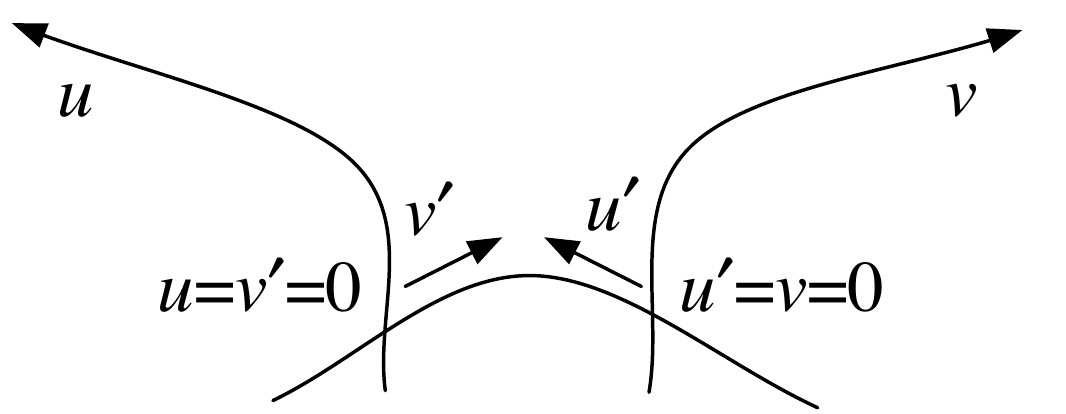}\\
\downarrow\\
\includegraphics[width=.3\textwidth]{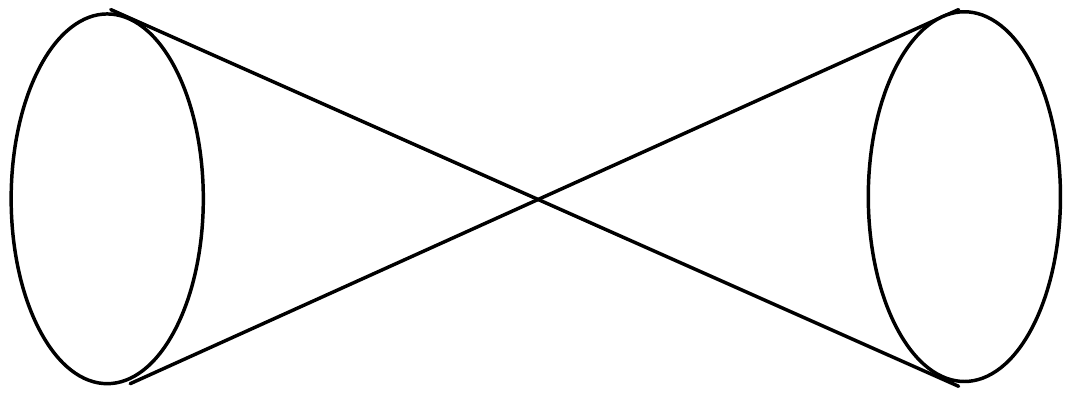}&
\includegraphics[width=.3\textwidth]{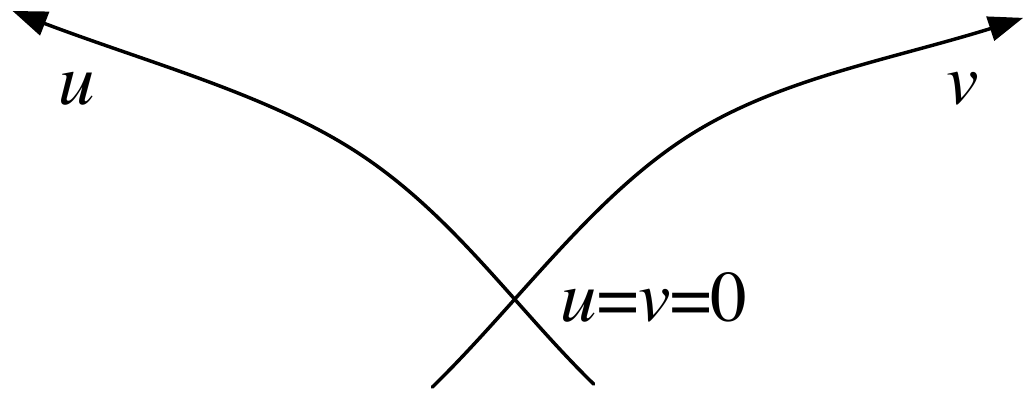}
\end{array}
\]
\caption{$\bR^4/\bZ_2$ のブローアップ。$u=v=0$ のところに、$u'=1/v'$ でパラメタ付けされた $S^2$ を挿入した。\label{blowup}}
\end{figure}

そこで、尖ったところに $S^2$ を差し込んで滑らかにします(図\ref{blowup})。
$S^2$ の北極では局所的には $(u,v')$ 但し $uv'=t$、
$S^2$ の南極では局所的には $(u',v)$ 但し $u'v=t$ が良い座標になっています。
$z$ の回転角が $(\epsilon_1+\epsilon_2)/2+a$、 
$w$ の回転角が $(\epsilon_1+\epsilon_2)/2-a$ でしたから、
北極での回転角は \begin{equation}
(\epsilon_1+\epsilon_2)+2a, -2a
\end{equation} 南極での回転角は \begin{equation}
2a, (\epsilon_1+\epsilon_2)-2a 
\end{equation}となります。そこで、Duistermaat-Heckman の公式をつかうと、\begin{equation}
\int e^{-2\pi H} d\mathrm{vol} = \frac{e^{-H_\text{北極}}}{((\epsilon_1+\epsilon_2)+2a)(-2a)}+
\frac{e^{-H_\text{南極}}}{(2a)((\epsilon_1+\epsilon_2)-2a)}
\end{equation} です。ブローアップするまえの $\bR^4/\bZ_2$ の積分を知るには、$S^2$ が小さくなる極限をとりますが、そうすると分子の $H$ は何にせよ原点に行ってゼロになりますので、\begin{align}
&\to \frac{1}{((\epsilon_1+\epsilon_2)+2a)(-2a)}+
\frac{1}{(2a)((\epsilon_1+\epsilon_2)-2a)}\\
&=\frac2{((\epsilon_1+\epsilon_2)+2a)((\epsilon_1+\epsilon_2)-2a)}\\
&=\frac12\frac1{((\epsilon_1+\epsilon_2)/2+a)((\epsilon_1+\epsilon_2)/2-a)}
\end{align}となり、先ほどの計算を再現しました。

$\cM_{2,2}$ 全体の積分にするには、$\bR^4$ の積分を掛ければよいです。
すると、\eqref{M21} では一項だったものを、\begin{equation}
Z_{2,1}= \frac{1}{\epsilon_1\epsilon_2}\frac{1}{((\epsilon_1+\epsilon_2)+2a)(-2a)}
+\frac{1}{\epsilon_1\epsilon_2}\frac{1}{(2a)((\epsilon_1+\epsilon_2)-2a)}\label{fubar}
\end{equation} と二項に分割したことになります。

\subsection{多重インスタントン計算}
以上前節では$\bR^4\times \bR^4/\bZ_2$上のガウス積分を非常にまわりくどく行ったわけですが、この方法の良いところは、前節で紹介した ADHM 構成と組み合わせると、一般の $N$, $k$ に対して計算が出来るところです\cite{Moore:1997dj,Nekrasov:2002qd,Flume:2002az,Nakajima:2003uh}。
導出は難しいのできちんとはできませんが、雰囲気だけ説明しましょう。$\SU(2)$ 1-インスタントンの場合、ブローアップは $uv=t^2$ の尖っているところに $S^2$ を埋め込みました。これは、ADHM 構成でいうと、尖っている状況は \begin{equation}
A_1 B_1+A_2B_2=0\hbox{、}\quad  \sum_{i=1,2} |A_i|^2 - |B^i|^2=0
\end{equation} で $(A_i,B_i)\mapsto (e^{i\theta} A_i,e^{-i\theta} B_i) $ という同一視をしていたものを、\begin{equation}
A_1 B_1+A_2B_2=0\hbox{、} \quad  \sum_{i=1,2} |A_i|^2 - |B^i|^2= r^2
\end{equation} と変えることに相当します。$B_i=0$ とすると、\begin{equation}
|A_1|^2+|A_2|^2=r^2 
\end{equation}という $S^3$ がありますが、これを $A_i\mapsto e^{i\theta}A_i$ で割ったのが、はめ込んだ $S^2$ になっていたわけです。$S^2$の北極南極は、\begin{equation}
(A_1,A_2)=(r,0)\hbox{、}\qquad 
(A_1,A_2)=(0,r)
\end{equation} にそれぞれ対応します。前者をとりますと、$\SU(2)$ 変換をすると勿論回ってしまいます \begin{equation}
(A_1,A_2)=(r,0) \mapsto ( e^{ia}r,0)
\end{equation}が、これはどうせ $(A_1,A_2)\simeq e^{i\theta}(A_1,A_2)$ と同一視するのでしたから、\begin{equation}
\simeq (r,0)
\end{equation} となって、$e^{i\theta}$ の範囲で固定されていることになります。$(A_1,A_2)=(0,r)$ の場合も同様ですね。しかし、$A_1,A_2$ の両方が $0$ でないと、$\SU(2)$ 回転すると \begin{equation}
(A_1,A_2)\mapsto (e^{ia} A_1,e^{-ia}A_2) 
\end{equation}と変換しますから、$e^{i\theta}$ で元に戻すことが出来ません。
ですから、肝は、$\SU(2)$ ゲージ回転および、$\bR^4$ の $(x_1,x_2)$ 平面と $(x_3,x_4)$平面の回転がどのように $e^{i\theta}$ 回転で相殺できるかということです。

一般には、\eqref{FOO}を保ったまま、\eqref{BAR} の右辺を $r^2$ にすることが $\cM_{N,k}$ を滑らかにすることになります。固定点は $k=1$ のときと同様、$B^i=0$ のところにあります。
固定点も同様に$\SU(N)$ ゲージ回転および、$\bR^4$ の $(x_1,x_2)$ 平面と $(x_3,x_4)$平面の回転がどのように $\U(k)$ 回転で相殺できるかで決まっています。
$\SU(N)$ ゲージ回転を $\diag(a_1,\ldots,a_N)$ で、$\bR^4$ の回転を $\epsilon_{1,2}$ ですることにすると、固定点はヤング図の $N$ 個組 $Y_1,\ldots,Y_N$ で、箱の数が合計 $k$ であるようなものでラベル付けがなされます。その際、相殺に用いる$\U(k)$ 回転は、対角成分が \begin{equation}
a_i + j\epsilon_1 + k \epsilon_2 
\end{equation} 但し $0\le j< (Y_i \text{の列の数})$, $0\le k<(Y_i \text{の $j$ 列目の高さ})$ となります。

\begin{figure}\[
\includegraphics[width=.2\textwidth]{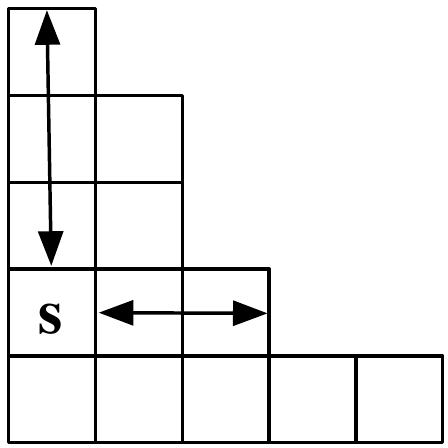}
\]
\caption{ヤング図 $Y$ の箱 $s$ に関する腕長 $A_Y(s)$、足長$L_Y(s)$。図の場合は $A_Y(s)=2$、
$L_Y(s)=3$ となる。$s$ が $Y$ の外にあるばあいは腕長、足長は負になる。\label{length}}
\end{figure}

ですから、$\cM_{N,k}$ をブローアップしたもの上での積分は、公式 \eqref{DH} による固定点 $p=(Y_1,\ldots,Y_N)$ 上の足し上げになります。元の$\cM_{N,k}$ 上での積分は、分子の $e^{-H}$ の項が全て $1$ になって、
\begin{equation}
Z_{N,k}=\int_{\cM_{N,k}}  e^{-2\pi(\epsilon_1 J_1+\epsilon_2 J_2+a_i K_i)} d\vol=
 \sum_{p=(Y_1,\ldots,Y_N)} \prod_{i} \frac{1}{\theta_{i,p}}
\end{equation} となります。各固定点での角度の計算 $\theta_{i,p}$ を実行すると、
\begin{multline}
Z_{N,k}= \sum_{Y_1,\ldots,Y_N} \prod_{i,j=1}^N
\prod_{s\in Y_i} (-L_{Y_j}(s)\epsilon_1+(A_{Y_i}(s)+1)\epsilon_2 + a_j-a_i )^{-1}\\
\times\prod_{t\in Y_j}((L_{Y_i}(t)+1)\epsilon_1-A_{Y_j}(s)\epsilon_2 + a_j-a_i )^{-1} \label{instantoncounting}
\end{multline}
という具体的な式で与えられます。
但し、$\sum a_i=0$とし、 $Y_1,\ldots,Y_N$ はヤング図で、箱の数が合計 $k$ 個であるようにします。$A_Y(s)$、$L_Y(s)$ はヤング図 $Y$ の箱 $s$ の腕長、足長と言われ、図 \ref{length} のように決めます。$\cM_{N,k}$ は $4Nk$ 次元ですから、分母には $2Nk$ 個回転角が並んでいるはずですので、それを確認してみてください。
$N=2$ の場合は $(a_1,a_2)=(a,-a)$ とすることにします。
$k=1$ とすると、$(Y_1,Y_2)=(\Young{1},0)$ か $=(0,\Young{1})$です。公式の組み合わせ論的な式を展開すると、\eqref{fubar} の二項をそれぞれ再現するのがわかると思います。

$k=2$ とすると、$(Y_1,Y_2)$ は $(\Young{2},0)$, 
$(\Young{11},0)$, $(\Young{1},\Young{1})$, 
$(0,\Young{2})$, $(0,\Young{11})$ の五通りあります。例えば $(\Young{2},0)$ からの寄与は \begin{equation}
\left[-4\epsilon_1^2\epsilon_2 (\epsilon_1-\epsilon_2) a (2a+\epsilon_1)(2a+\epsilon_1+\epsilon_2) (2a+2\epsilon_1+\epsilon_2) \right]^{-1}
\end{equation} となります。他の項も頑張って計算しますと、\begin{equation}
Z_{2,2}=\frac{( 8(\epsilon_1+\epsilon_2)^2 +\epsilon_1\epsilon_2 - 8a^2)}
{\epsilon_1^2\epsilon_2^2 ((\epsilon_1+\epsilon_2)^2-4a^2)
((2\epsilon_1+\epsilon_2)^2-4a^2) ((\epsilon_1+2\epsilon_2)^2-4a^2) }\label{Z2}
\end{equation} となります。

なんだかややこしい結果になりましたが、もうすこしだけ物理的内容をとりだしてみましょう。インスタントン一つの場合は \eqref{fubar} でした。インスタントンふたつが遠く離れていれば、1-インスタントン解を二つ重ね合わせれば2-インスタントン解ができます。ですから、
モジュライ空間の外部では、おおよそ $\cM_{2,2} \sim (\cM_{2,1})^2/\bZ_2 $ でした。
ですから、$\cM_{2,2}$上の積分は、おおよそ$\cM_{2,1}$上の積分のほぼ二乗で与えられて、差はインスタントン二つが互いに近づいた際の相互作用からくるはずです。それを実際に確かめましょう。

$|a|\gg \epsilon_{1,2}$ としたとします。すると、
\eqref{Z2} にある $Z_{2,2}$  も \eqref{fubar}$^2$ で与えられる $Z_{2,1}{}^2$もそれぞれ $(\epsilon_1\epsilon_2)^{-2}$ の因子があります、すなわち時空の箱のサイズの二乗の因子があります。これは、時空をふたつ物体が動いている積分から来ます。一方、相互作用の効果は \begin{equation}
Z_{2,2}-\frac12 Z_{2,1}{}^2=\frac{20a^2+7\epsilon_1^2+16\epsilon_1\epsilon_2 + 7\epsilon_2^2}{\epsilon_1\epsilon_2((\epsilon_1+\epsilon_2)^2-4a^2)((2\epsilon_1+\epsilon_2)^2-4a^2)((\epsilon_1+2\epsilon_2)^2-4a^2)} \label{hoge}
\end{equation}となって、 $(\epsilon_1\epsilon_2)^{-1}$ の項が相殺し、 $(\epsilon_1\epsilon_2)^{-1}$ の寄与しかありません。これは、時空の各点を相互作用点として、そこに二つのインスタントンが近づいてきたときにおこる寄与がある、ということになっています。別の言い方をしますと、\begin{equation}
Z^\text{instanton}_{\epsilon_1,\epsilon_2;a} = 1 + q Z_{2,1} + q^2 Z_{2,2} +\cdots 
\end{equation}としますと、 $q^k$ の項はインスタントンが $k$ 個ありますから $(\epsilon_1\epsilon_2)^{-k}$ の因子がありますが、対数をとってやると \begin{align}
\log Z^\text{instanton}_{\epsilon_1,\epsilon_2;a}&= \frac{q}{\epsilon_1\epsilon_2}
\frac12\frac1{((\epsilon_1+\epsilon_2)/2+a)((\epsilon_1+\epsilon_2)/2-a)}  \\
&+\frac{q^2}{\epsilon_1\epsilon_2}\frac{20a^2+7\epsilon_1^2+16\epsilon_1\epsilon_2 + 7\epsilon_2^2}{((\epsilon_1+\epsilon_2)^2-4a^2)((2\epsilon_1+\epsilon_2)^2-4a^2)((\epsilon_1+2\epsilon_2)^2-4a^2)}+\cdots \\
&\to \frac{q}{\epsilon_1\epsilon_2} \frac{-1}{a^2} + \frac{q^2}{\epsilon_1\epsilon_2} \frac{-5}{16a^4} + \cdots \label{bosh}
\end{align} となって、$q$ のべきにかかわらず時空体積の因子 $(\epsilon_1\epsilon_2)^{-1}$ が毎回でることになります。これは、$k$ インスタントンが相互作用するのは各時空点においてである、という様子を捉えているわけです。\eqref{bosh} では $|a|\gg \epsilon_{1,2}$ という極限をとりました。

\section{二次元と四次元の関係}
\subsection{対応関係}
さて、もうそろそろはじめに何をやったのかお忘れではないかと思いますが、前々節では、二次元共形場理論で、状態 $\ket{\Delta}$ から生成される Verma 表現を考えて、その中でコヒーレント状態 \begin{equation}
L_1\ket{\Delta,\lambda}=\lambda \ket{\Delta,\lambda}\hbox{、}\qquad
L_2\ket{\Delta,\lambda}=0
\end{equation}を取り、\begin{equation}
\vev{\Delta,\lambda|\Delta,\lambda}=\vev{\Delta,1| \lambda^{2(L_0-\Delta)} |\Delta,1}=1+\frac{\lambda^2}{2\Delta}+\frac{\lambda^4(c+8\Delta)}{4\Delta((1+\Delta)c-10\Delta+16\Delta^2)}+\cdots \label{A}
\end{equation}を計算しました。

また前節では、インスタントンの統計力学を考えました。時空を超立方体の箱にいれる替わりに、時空の回転およびゲージの回転に対して化学ポテンシャルを入れ、\begin{equation}
Z^\text{instanton}_{\epsilon_1,\epsilon_2,a}=\sum_k q^k \int_{\cM_{2,k}}  e^{-2\pi(\epsilon_1 J_1 + \epsilon_2 J_2 + a K)} d\vol \hbox{。}
\end{equation}を考えました。但し $N=2$ の $\SU(2)$ の場合を考えます。これを頑張って計算すると、\begin{multline}
=1+ \frac{q}{\epsilon_1\epsilon_2} \frac{2}{(\epsilon_1+\epsilon_2)^2 -4a^2}  \\
+ \frac{q^2}{\epsilon_1^2\epsilon_2^2} \frac{( 8(\epsilon_1+\epsilon_2)^2 +\epsilon_1\epsilon_2 - 8a^2)}
{ ((\epsilon_1+\epsilon_2)^2-4a^2)
((2\epsilon_1+\epsilon_2)^2-4a^2) ((\epsilon_1+2\epsilon_2)^2-4a^2) } + \cdots\label{B}
\end{multline} となりました。

ふたつの結果\eqref{A}, \eqref{B} を見比べますと、第二項までは \begin{equation}
\lambda^2=\frac{q}{(\epsilon_1\epsilon_2)^2}\hbox{、} \qquad
\Delta=\frac1{\epsilon_1\epsilon_2}(\frac{(\epsilon_1+\epsilon_2)^2}{4} -a^2) \label{co}
\end{equation} とすれば合致します。第三項は、さらに \begin{equation}
c=1+6\frac{(\epsilon_1+\epsilon_2)^2}{\epsilon_1\epsilon_2}
\end{equation}とすると合います。ここまでははじめの三項を使って3つの変数を関係付けただけのように見えるかもしれませんが、第四項、第五項とどんどん計算すると、上記の変数の対応で \eqref{A} と \eqref{B} がいつまでも合致するということがわかります\cite{Alday:2009aq,Gaiotto:2009ma}。前節で説明した \eqref{A} と \eqref{B} を計算するアルゴリズムは前節と前々節に書きましたので、是非次の項を計算してみてください。 手計算でやると 3-インスタントンのあたりから突然面倒になりますので、プログラムを書いたほうがいいでしょう。Mathematica で実装したものが
プレプリントページの ancillary files のところに置いてありますので、それをご覧ください。

量 $\vev{\Delta,\lambda|\Delta,\lambda}$ は二次元の物理から出て来たものです。量 $Z^\text{instanton}_{\epsilon_1,\epsilon_2,a}$ は四次元の物理から出て来たものです。これらが一致するという現象を、どのように理解すれば良いでしょうか。
双方とも計算方法は判っています。コヒーレント状態の長さの二乗は公式 \eqref{russian} で具体的に与えられていますし、インスタントンの分配関数は公式 \eqref{instantoncounting} で具体的に与えられています。
ですから、背景は忘れて、これらの公式を睨んで証明しよう、とすることが出来、実際そのような証明が既に知られています\cite{Poghossian:2009mk,Fateev:2009aw,Hadasz:2010xp}。
また、モジュライ空間に作用する無限次元代数を構成するというのは、数学では幾何学的表現論という一分野をなすほどです。たとえば、$\bC^2/\Gamma$ 但し $\Gamma\subset \SU(2)$ という空間上でのインスタントンのモジュライ空間を考えると、$\Gamma$ 型の Kac-Moody 代数の表現があらわれるというのが \cite{Nakajima} で示されています。
ですから、上記等式は$\Gamma$ で割る替わりに $\epsilon_{1,2}$ を入れた自然な拡張になっていて、幾何学的表現論からの証明がそろそろ発表されるという噂です\cite{MO}。
これら二つの手法は、図\ref{relation} で言えば、数学の中に留まって関係を理解しよう、というものです。
しかし、僕の専門は数学ではなく弦理論ですので、厳密ではない弦理論の世界を通過して、どのようにこの関係が理解できるか、というのを説明したいと思います。

\begin{figure}
\[
\includegraphics[width=.6\textwidth]{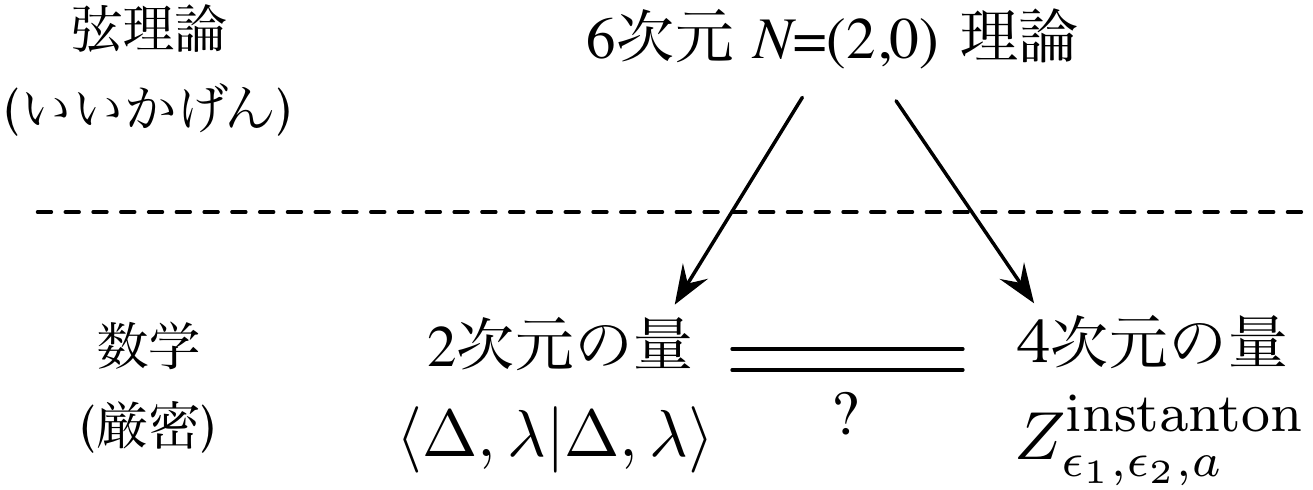}
\]
\caption{二次元の量と四次元の量との関係\label{relation}}
\end{figure}

\subsection{まず五次元へ}
我々の理解したい式は \begin{equation}
\vev{\Delta,1|\lambda^{2(L_0-\Delta_0)}|\Delta,1} = \sum_k q^k \int_{\cM_{2,k}} e^{-2\pi(\epsilon_1 J_1 +\epsilon_2 J_2 + aK)} d\vol
\end{equation}というものでした。二次元と四次元の量を比較しているという以前に、
左辺は波動関数の二乗の形をしており、右辺は単なる積分です。これをどう比較すればよいでしょう? そのために、右辺を以下のように考え直してみましょう。$\psi_k(t)$ を、$\cM_{2,k}$ 上の関数で \begin{equation}
\psi_k(t)=e^{-\pi(\epsilon_1 J_1+\epsilon_2 J_2 + aK)}
\end{equation} で定めます。具体的には、$k=1$ では \begin{equation}
\psi_1(x_1,x_2,x_3,x_4,z,w)=e^{-\pi(\epsilon_1(x_1^2+x_2^2)+\epsilon_2(x_3^2+x_4^2) + 
(\frac{\epsilon_1+\epsilon_2}2+a)|z|^2+(\frac{\epsilon_1+\epsilon_2}2-a)|w|^2 )}
\end{equation}で、ガウス型の関数です。すると、右辺は \begin{equation}
=\sum_k q^k \int_{\cM_{2,k}} |\psi_k(t)|^2 d\vol
\end{equation}です。
$\phi_k(t)$を、$\cM_{2,k}$ 上を動いている量子力学的粒子の波動関数だと思って $\ket{\phi_k}$ と書きましょう: \begin{equation}
\ket{\phi_k} \in \cH(\cM_{2,k})
\end{equation} ただし $\cH(\cM_{2,k})$ は $\cM_{2,k}$ 上の波動関数のなすヒルベルト空間です。すると、右辺はさらに \begin{equation}
=\sum_k q^k \vev{\phi_k|\phi_k} 
\end{equation}と書けます。そこで、さらに \begin{equation}
\ket{\phi}=\ket{\phi_0}\oplus \ket{\phi_1} \oplus \cdots 
\in \cH(\cM_{2,0}) \oplus \cH(\cM_{2,1}) \oplus \cdots \label{phi}
\end{equation} というベクトルを考え、ハミルトニアン $H$ を $\cH(\cM_{2,k})$ で固有値 $k$ をもつような演算子だとすると、関係式は\begin{equation}
\vev{\Delta,1|\lambda^{2L_0-\Delta}|\Delta,1}=\vev{\phi | q^H | \phi}
\end{equation}となります。$\ket{\Delta,1}$ も \begin{equation}
\ket{\Delta,1}=\ket{\psi_0} \oplus \ket{\psi_1} \oplus \ket{\psi_2} \oplus \cdots 
\in V_0 \oplus V_1 \oplus V_2 \oplus \cdots
\end{equation} という展開がありました、但し $V_k$ は次数が $k$ の成分で、そこでは $L_0-\Delta=k$ なのでした。

$V_k$ は有限次元でした: $\dim V_k = p_k$ ただし $p_k$ は $k$ を正整数の和として書く方法の数。一方で $\cH(\cM_{2,k})$ は当然無限次元です。しかし、何らかの意味で \begin{equation}
V_k \subset \cH(\cM_{2,k}) 
\end{equation}と自然に埋め込まれており、その埋め込みのもとで \begin{equation}
\ket{\psi_k} = \ket{\phi_k}
\end{equation} となっているならば、我々の関係式は自然に従います。

しかし、インスタントンのモジュライ空間を動く粒子というのはどういうことでしょうか? これは、四次元のゲージ理論でなく、五次元のゲージ理論を考えると自然に現れます。
ヤンミルズ場を $\bR^4$ でなく $\bR^5 =\bR^4\times \bR_t $ 上で考えましょう。付け加えた一方向を時間だと思うことにします: $(x_1,x_2,x_3,x_4,t) \in \bR^5$\hbox{。}  作用は
\begin{equation}
\int \frac1{2g_\text{5d}^2} \tr F_{\mu\nu} F_{\mu\nu} dt d^4x  \label{5daction}
\end{equation}です。
五次元の配位をひとつ $\cA_\mu(x_1,x_2,x_3,x_4,t)$ をとると、各時刻 $t$ 毎に $\bR^4$ 上の配位 \begin{equation}
A_\mu(x_i;t) = \cA_\mu(x_1,x_2,x_3,x_4,t)
\end{equation} が定まっていると思えます。時刻 $t=t_1$ での $A_\mu(x_i;t_1)$インスタントン数を $k$ とすると、インスタントン数は整数ですから、他のいつの時刻 $t=t_2$ でもインスタントン数は $k$ になります。
量子力学にするために経路積分をすることを考えますと、各時刻でエネルギーを極小にするために、各時刻 $t$ で $A_\mu(x_i;t)$ が反自己双対であるような配位がもっとも寄与が大きくなります(図\ref{5d})。
各時刻でのエネルギーは \begin{equation}
\int \frac1{2g_\text{5d}^2} \tr F_{\mu\nu} F_{\mu\nu} d^4x  \sim \frac{8\pi^2 k}{g_\text{5d}^2}
\end{equation}となります。エネルギーと質量は等価ですから、
インスタントン一つが質量 $8\pi^2/g_\text{5d}^2$ の粒子に見えるわけです。

\begin{figure}\[
\includegraphics[width=.6\textwidth]{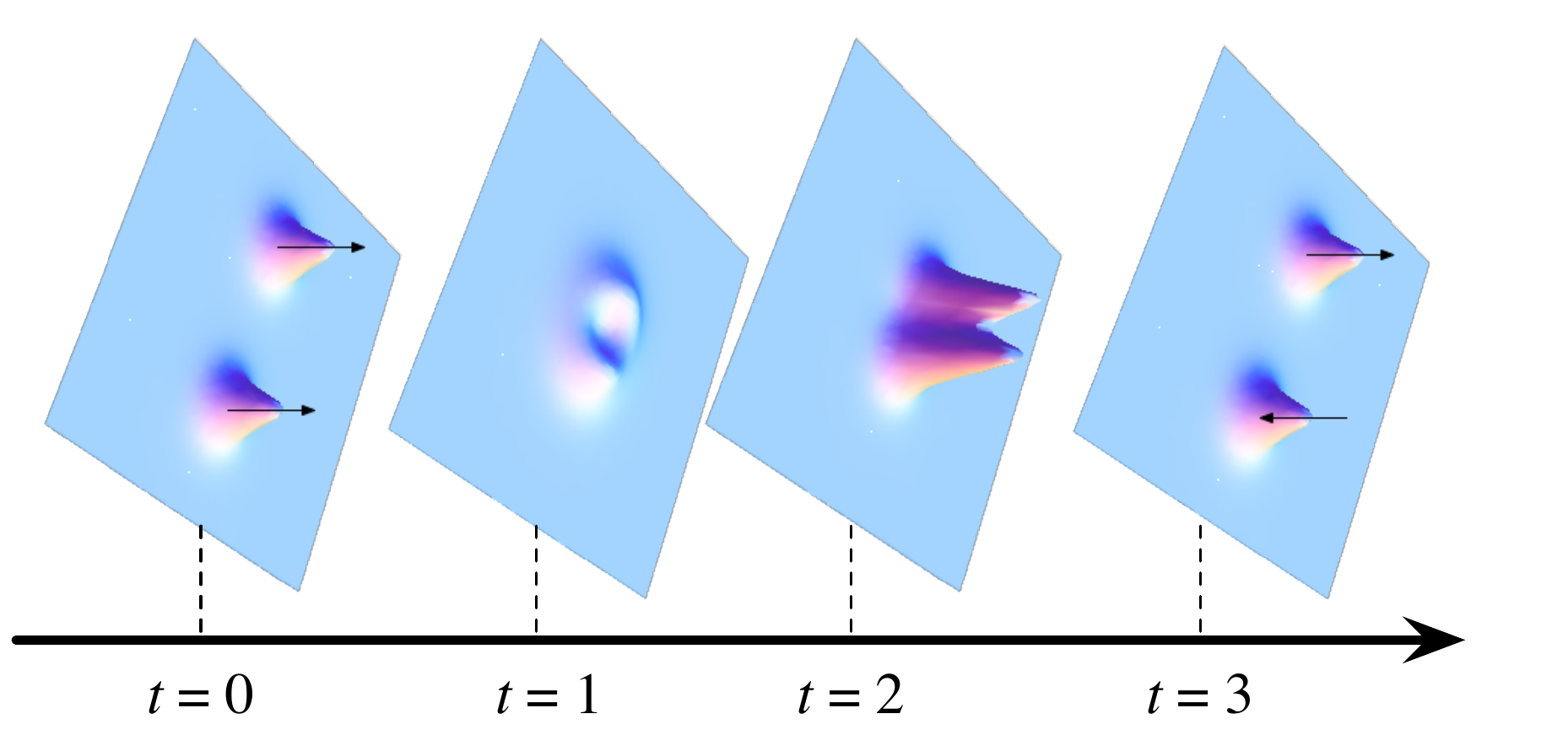}
\]
\caption{五次元ゲージ理論と反自己双対解の変化\label{5d}}
\end{figure}

さて、勝手なインスタントン数 $k$ の反自己双対解は $4Nk$ 個のパラメタ $s_i$ で決まっているはずですから、時刻 $t$ に依存して $s_i(t)$ が決まりました: \begin{equation}
s_i(t): \bR_t \to \cM_{N,k}\hbox{。}
\end{equation} ですから、五次元のゲージ理論で、経路積分に最も寄与の大きい部分は、インスタントンのモジュライ空間を動く量子力学的粒子の運動で捉えられることがわかりました。

しかし、我々は四次元のゲージ理論を考えていました。五次元のゲージ理論から四次元のゲージ理論をつくる簡単な方法は、一つの方向を「コンパクト化」することです。例えば、$t$ 方向を $[0,L]$ の線分にしてしまいましょう。すると、理論を $L$ より非常におおきなスケールでみる限りは、$t$ 方向を区別することはできず、実質四次元の理論になります。非常に安直には、五次元の作用\eqref{5daction} において、$F_{\mu\nu}$ が $t$ 方向に変化しなければ、$dt$ 積分をしてしまって \begin{equation}
\int \frac1{2g_\text{4d}^2} \tr F_{\mu\nu} F_{\mu\nu} d^4x 
\end{equation} とできるというわけです。但し、\begin{equation}
\frac1{g_\text{4d}^2} = \frac{L}{g_\text{5d}^2}\hbox{。} \label{45}
\end{equation}

一般に、このようなコンパクト化をすると Kaluza-Klein 粒子というものが現れます。五次元の時空に質量のない粒子があったとしましょう。すると、エネルギーと運動量は \begin{equation}
E^2 = \vec{p}\,^2 + p_5^2
\end{equation} を満たします、ただし $\vec p$ は $\bR^3$ 方向の運動量で、$p_5$ は第五方向の運動量とします。第五方向を $L$ にコンパクト化すると、量子力学的には運動量は波動関数の位相 $e^{ i \vec p\cdot \vec x}$ ですから、$2\pi p_5 L$ は整数でないといけません。それを $k$ とすると、$p_5 = 2\pi k /L$ となります。すると、四次元の立場からは、\begin{equation}
E^2 - \vec{p}\,^2= \left(\frac{2\pi k}{L}\right)^2
\end{equation}となって、質量が $2\pi k/L$ の粒子が現れます。これは $L$ が小さければどんどん重くなるので、$L$ が小さいほど測りにくくなるわけです。

\begin{figure}
\[
\includegraphics[width=.8\textwidth]{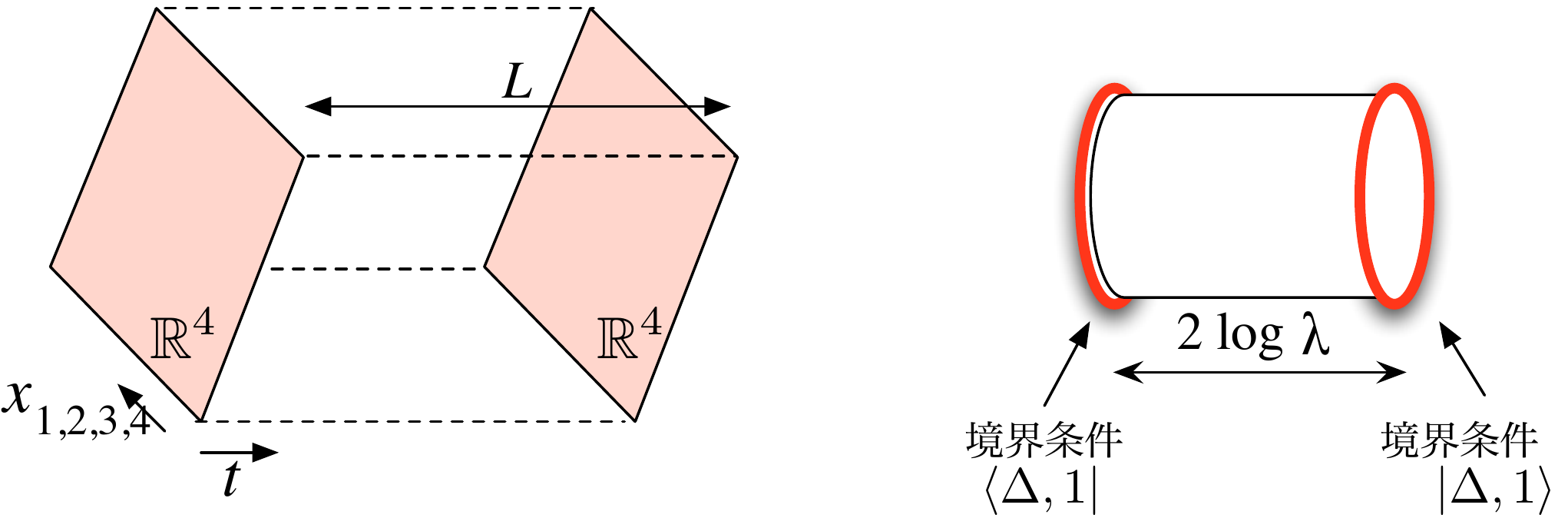}
\]
\caption{左: 五次元ゲージ理論から四次元ゲージ理論をつくる。右: 図\ref{HH}を再掲。二次元共形場理論のコヒーレント状態のノルム。\label{interval}}
\end{figure}

ゲージ理論に話を戻しますと、我々は四次元の結合定数が $g_{4d}$ のゲージ理論を考えたいので、五次元のゲージ理論を \eqref{45} で定まる長さ $L$ の線分にコンパクト化します 図\ref{interval}。
すると、この系の分配関数は、\begin{equation}
Z=\vev{\Phi |e^{-LH} |\Phi}
\end{equation} で与えられることになります。但し $\Phi$ は $t=0$ および $t=L$ での境界条件から定まる状態で、$H$ は $t$ 方向への時間発展の演算子です。

\subsection{さらに六次元へ}
ここまでは非常に一般的な考察でしたが、五次元のゲージ理論として、単にゲージ場だけの理論でなく、最大超対称 $\SU(N)$ ゲージ理論とよばれるものを取り、境界条件で半分の超対称を保つものを使うと、四次元の理論として純 $\cN=2$ $\SU(N)$ゲージ理論というものになります。この理論に更に Nekrasov の $\epsilon_{1,2}$ 変形というものを加えると、分配関数が我々の扱ってきた統計力学模型と一致することが知られています。いま我々は $N=2$ を考えています。すると、上記 $\ket{\Phi}$ は $\ket{\phi}$ と一致します。
さて、これは更に共形場理論のコヒーレント状態 $\ket{\Delta,1}$ と一致するのですが、状況を比べると五次元を線分において四次元理論をつくるのと、
コヒーレント状態のノルムを計算する状況はほとんど同じですね(図\ref{interval})。
五次元方向の長さ $L$ は\begin{equation}
L\sim \frac{1}{g_\text{4d}^2} \sim \log q 
\end{equation}でしたが、円柱の横幅は $\sim\log \lambda$ で、\eqref{co} で見た対応関係から期待される通りです。

ですから、五次元最大超対称 $\SU(2)$ ゲージ理論の $t$ 方向の発展の演算子 $H$ は、二次元の理論の演算子 $L_0$ と同一視すべきです。同じことですが、図 \ref{interval} の第五方向、すなわち $L$ 方向と、図 \ref{HH} の円柱の $\log \lambda$ 方向は同一視すべきです。
五次元の理論の五方向はすべて $\SO(5)$ 回転で等価です: \begin{equation}
\text{$\bR^4$の一方向} \xrightarrow{\text{$\SO(5)$ 回転}} \text{第五方向}
\end{equation}
一方で、共形変換は $L_0$ だけでなく、$L_{n}$ を含み、特に $L_{\pm 1}$ は円柱の $\bR$ 方向を円柱の$S^1$方向に回す変換を含んでいます。物理的には、円柱の中の二方向は等価なわけです。\begin{equation}
\text{円柱の $\bR$ 方向} \xrightarrow{\text{$L_1$ 回転}} \text{円柱の$S^1$方向}
\end{equation} しかし、\begin{equation}
\text{五次元理論の第五方向} = \text{円柱の $\bR$ 方向} 
\end{equation}です。これらを組み合わせると、五次元理論の $\bR^4$ 方向は、円柱の空間方向へ回転させることが出来るという主張に至ります。そのためには、六次元の理論が必要です。
純粋に五次元最大超対称 $\SU(2)$ ゲージ理論を考えているつもりだったが、それは六次元のある理論を一周 $R$ の円周にコンパクト化したものであると思うべきである、ということです。

六次元の理論を円周にコンパクト化したのであれば、すぐ前におさらいしたように、整数 $k$ に対して重さ $2\pi k/R$ の Kaluza-Klein 粒子がでるはずですが、確かに、五次元理論にはインスタントン粒子があり、その重さは $8\pi^2 k/g_\text{5d}$ でした。ですから、五次元理論のインスタントン粒子は、実は六次元理論の Kaluza-Klein 粒子であり、 \begin{equation}
R= \frac{g_\text{5d}^2}{4\pi}  \label{56}
\end{equation} である、ということがわかります。

ですから、我々は六次元の理論を、$S^1\times [0,L] \times \bR^4$ 上で考えていたのです。この系を解析する際に、
\begin{itemize}
\item ここで、$S^1$ が非常に小さいとすると、$[0,L]\times \bR^4$ 上で五次元ゲージ理論を、さらに $L$ も小さいとして、結局 $\bR^4$ で四次元ゲージ理論を考えるという方法があります。
\item 一方、 $\bR^4$ 方向にはポテンシャル $\epsilon_1(x_1^2+x_2^2) + \epsilon_2(x_3^2+x_4^2)$ を入れているので、$\epsilon_{1,2}$ が大きいと、$\bR^4$ 方向の箱を非常に小さくすることが出来ます。すると、$S^1\times [0,L]$ 上で二次元理論を考える、ということになります。
\end{itemize}
この二通りの評価法を比較することにより、\begin{equation}
Z^\text{instanton}_{\epsilon_1,\epsilon_2,a} = \vev{\Delta,\lambda|\Delta,\lambda}
\end{equation} が得られるわけです。

では、この六次元理論はなんでしょうか? 先ほど、五次元ゲージ理論を長さ $L$ の線分にコンパクト化すれば、おおよそ四次元ゲージ理論になることを説明しましたから、同様に、六次元ゲージ理論を一周 $R$ の円周にコンパクト化しているのではないか、と安直には思えます。しかし、そうすると \eqref{45} と同様にして、 \begin{equation}
\frac{1}{g_\text{5d}^2 } \propto R 
\end{equation}となってしまい、\eqref{56} とはまったく逆になってしまいます。

きちんと \eqref{56} を出すような六次元理論は、我々の業界では「六次元 $\cN=(2,0)$ 理論」と呼ばれています。これの性質ははっきりとはわかっていませんが、ゲージ理論の $F_{\mu\nu}=\partial_\mu A_{\nu} -\partial_\nu A_\mu +[A_\mu,A_\nu] $ の替わりに、$F_{\mu\nu\rho}=\partial_\mu B_{\nu\rho} + \cdots $ という場があり、\begin{equation}
F_{\mu\nu\rho}=\tilde F_{\mu\nu\rho} \quad \text{但し} \quad \tilde F_{\mu\nu\rho}=
\frac16 \epsilon_{\mu\nu\rho\alpha\beta\gamma} F_{\alpha\beta\gamma}
\end{equation} が成り立っている「ようなもの」だと思われています。この六次元理論はいろいろな次元での超対称ゲージ理論の親玉だと思われており、最近活発に研究されています (数学者向けのまとめは \cite{Witten:2009at,Witten:2009mh} 等を参照のこと)。

五次元の最大超対称ゲージ理論は、結合定数 $g_\text{5d}$ を大きくすると、インスタントン粒子が軽くなり、一周 $R\sim g_\text{5d}^2$ の円周を生成して六次元の理論になってしまうわけですが、この事実は \cite{Dijkgraaf:1997vv} に示唆されて 1997 年に \cite{Berkooz:1997cq} がはじめに指摘しました。
これは 1995 年から理解され始めた「超弦理論は M 理論である」という事実の一環です。
実際、Type IIA 超弦理論は10次元の理論なのですが、弦理論の結合定数を大きくすると、D0-ブレーンという粒子が軽くなり、円周を生成して11次元の理論になってしまいます。
Type IIA 超弦には D4-ブレーンという五次元にひろがった物体があり、この操作にともなって、M理論の M5-ブレーンという6次元にひろがった物体が生成された円周に巻き付いている状態になります(このあたりの詳細は、教科書 \cite{Polchinski} 等を参照のこと)。
D4-ブレーンが $N$ 枚重なっていると、その上には五次元最大超対称 $\SU(N)$ ゲージ理論が住みます。これを強結合にすると、M5-ブレーンが $N$ 枚重なっていることになります。この上に住んでいるのが、我々の知りたい6次元$\SU(N)$型 $\cN=(2,0)$ 理論です。

この六次元理論ははっきりとはわかっていませんが、これを一周 $R$ の円周上にコンパクト化し、さらにこれを長さ $L$ の線分にコンパクト化した系の分配関数を二通りの方法で計算しようとしたところ、僕と共著者はひとつの方法では $\vev{\Delta,\lambda|\Delta,\lambda}$ を、もうひとつの方法では $Z^\text{instanton}_{\epsilon_1,\epsilon_2,a}$ になることが分かったので、これらは等しいはずだ、という事実に辿り着いたのでした \cite{Alday:2009aq,Gaiotto:2009ma}\footnote{正直なことを言いますと僕が事実に辿り着いたというのは言い過ぎです。僕はインスタントンの分配関数の計算を修士論文にしていたのですが、それを知っていた D. Gaiotto がある日、 F. Alday との共同研究の過程で、$Z^\text{instanton}$ が二次元共形場理論のこういう量で掛ける筈だとなったんだが確かめてくれないか、と僕に言いました。そこで、数年前に書いた Mathematica プログラムをパソコンから掘り出してわけもわからず計算してみると確かに一致していたので驚愕した、というのが \cite{Alday:2009aq} の真相です。その後、さらに例を簡単化したのが \cite{Gaiotto:2009ma} で、この講義では簡単化したものだけを説明しました。}。

\subsection{拡張}

さて、ここまでは \begin{equation}
\vev{\Delta,\lambda| \Delta,\lambda} = Z^\text{instanton}_{\epsilon_1,\epsilon_2,a}
\end{equation} というただ一つの関係式に絞ってなるべく具体的に説明をしてきたつもりですが、
前節での説明から、すこし弦理論の設定をかえると、幾らでも関連する関係式が得られることがわかるでしょう。それをいくつか述べておしまいにしたいと思います。

まず、この等式では左辺に二次元のビラソロ代数のコヒーレント状態があり、
右辺では $\SU(2)$ のインスタントンの分配関数があります。
$\SU(2)$ のかわりに $\SU(N)$ にするとどうなるでしょうか? 
右辺を計算することは簡単で、公式は既に \eqref{instantoncounting} に書きました。
左辺はどうすればよいでしょう? 
1980年代遅くから1990年代初頭にかけて、ビラソロ代数は $W_N$ 代数とよばれる無限次元代数のクラスの中で $N=2$ の一番簡単なもの、すなわち $W_2$ 代数であるということが認識されました。(W代数については、レビュー\cite{Bouwknegt:1992wg} や、「数理物理2004」の脇本先生の講義録 \cite{Wakimoto} 等を参照。)
ですから、上式を一般化して \begin{equation}
\text{$W_N$ 代数のコヒーレント状態のノルム} = \text{$\SU(N)$インスタントンの分配関数}
\end{equation}ということを考えるのは自然です\cite{Wyllard:2009hg,Mironov:2009by}。例えば、$N=3$ の場合は、$L_n$ に加えて $W_n$ という演算子があり、交換関係は
\begin{align}
&[L_n, L_m]=(n-m)L_{n+m}+\frac{c}{12}(n^3-n)\delta_{n,-m}\hbox{、}\\
&[L_n, W_m]=(2n-m)W_{n+m}\hbox{、}\\
&[W_n, W_m]=\frac{1}{48}\Big[ \frac{c(22+5c)}{3\cdot 5!}n(n^2-1)(n^2-4)\delta_{n,-m}
+{16}(n-m)\Lambda_{n+m}\nonumber\\
&\qquad\qquad\qquad +(22+5c)(n-m) \left( \frac{(n+m+2)(n+m+3)}{15}-\frac{(n+2)(m+2)}{6} \right)L_{n+m}\Big]\hbox{。}
\end{align}で与えられます、但し
$\Lambda_n$ は
\begin{align}
\Lambda_n=\sum_{m\le -2}L_m L_{n-m}+\sum_{m\ge -1} L_{n-m}L_m-
\frac3{10}(n+2)(n+3)L_n
\end{align}
で定義されます。ここで、$W_m$ の規格化は通常のものの $\sqrt{(22+5c)/216}$倍にしました。自由場表示からはこちらのほうが自然です。$W_3$代数の基本的な表現は \begin{equation}
L_0\ket{\Delta,w}=\Delta\ket{\Delta,w}, \ 
W_0\ket{\Delta,w}=w\ket{\Delta,w} 
\end{equation} で $n>0$ なら$L_n$, $W_n$ を掛けると消えるような状態から生成されます。それらの線形結合から、コヒーレント状態を \begin{equation}
W_1\ket{\Delta,w,\lambda}=\lambda \ket{\Delta,w,\lambda}
\end{equation} で定めると、そのノルムが式\eqref{instantoncounting}で $N=3$ とした $Z^\text{instanton}$ と一致することが知られています \cite{Mironov:2009by,Taki:2009zd}。ここで、パラメタの対応は \begin{equation}
c=2+24(b+\frac1b)^2,\quad
\Delta=(b+\frac1b)^2-a_1^2-a_1a_2-a_2^2,\quad
w=ia_1a_2(a_1+a_2)
\end{equation} とします。但し、簡単のため $\epsilon_1=b$, $\epsilon_2=1/b$ としさらに$(a_1,a_2,a_3)=(a_1,a_2,-a_1-a_2)$ と取りました。

\begin{figure}\[
\includegraphics[width=.7\textwidth]{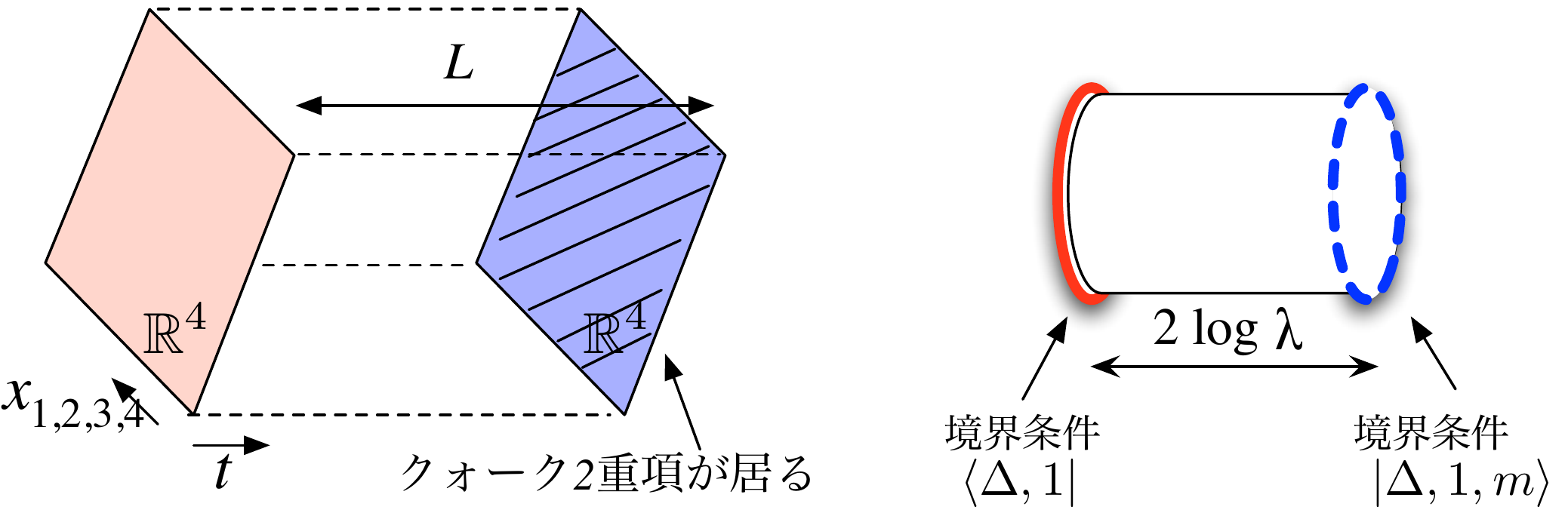}
\]
\caption{2重項クォークを一つ入れた状況。左: 五次元のゲージ理論を線分に置いた。右: 対応する二次元共形場理論の状況。\label{matter}}
\end{figure}

また、$\SU(2)$ のままで、ゲージ理論に物質場を足すことも出来ます。例えば、現実の「弱い力」のもとではクォークは $\SU(2)$ の二重項ですから、それに倣って二重項のクォークの寄与をインスタントン分配関数に加えることができます。すると、\eqref{instantoncounting} の分子にもいろいろ項が加わることが知られています。我々の設定でクォークをひとつ足す簡単な方法は、片方の境界にだけクォークを足すことです。すると、対応して二次元の共形場理論では、片側の境界条件が変更されます(図\ref{matter})。すると、\begin{equation}
\vev{\Delta,\lambda|\Delta,\lambda,m} = Z^\text{instanton}_\text{with quark} 
\end{equation}という等式が成り立ちます\cite{Gaiotto:2009ma}。ただし、$\ket{\Delta,\lambda,m}$ は $L_1$ も $L_2$ もノンゼロの固有値をもつようなコヒーレント状態です: \begin{equation}
L_1\ket{\Delta,\lambda,m}=\lambda\ket{\Delta,\lambda,m}\hbox{、}\qquad
L_2\ket{\Delta,\lambda,m}=\sqrt{\lambda} m\ket{\Delta,\lambda,m}\hbox{。}
\end{equation}
はじめの論文 \cite{Alday:2009aq} で扱われたのは、さらに複雑にクォーク二重項を四つ加えた場合でした。

別の拡張として、$W_N$ 代数はアファイン $\SU(N)$ 代数から量子 Drinfeld-Sokolov 還元という手法で作ることが出来ますが、還元にはデータとして $N$ の分割 $\rho=(N_1,\ldots,N_k)$ を指定してやることができます。$W_N$ は特に $\rho=(N)$ という場合ですが、一般の $\rho$ に対して $W(\SU(N),\rho)$ 代数というものがあります。
これを出すにはゲージ理論側にどのような変更を加えれば良いかというのも知られており、$\bR^4$ 上のインスタントンを考える際に、$x_1=x_2=0$ の平面に沿ってゲージ場に特異性 
\begin{equation}
A_\mu dx^\mu \sim \mathrm{diag}(
\underbrace{\alpha_{(1)},\ldots,\alpha_{(1)}}_{\text{$n_1$ times}},
\underbrace{\alpha_{(2)},\ldots,\alpha_{(2)}}_{\text{$n_2$ times}},
\ldots,
\underbrace{\alpha_{(k)},\ldots,\alpha_{(k)}}_{\text{$n_k$ times}} ) i d\theta, 
\end{equation} を入れれば良いです\cite{Braverman:2010ef,Wyllard:2010vi,Kanno:2011fw}。

また、$\bR^4$ 上でばかりインスタントンを考えてきましたが、そのかわりに $\bR^4/\Gamma$ 上でインスタントンを考えればどうなるでしょうか。ただし、$\Gamma\subset \SU(2)$ とします。$\epsilon_{1,2}$ を入れない状況は \cite{Nakajima} によって調べられ、$\Gamma$ 型のKac-Moody 代数がでることが知られていました。$\Gamma=\bZ_m$ の場合はさらに $\epsilon_1$, $\epsilon_2$ による変形を入れることが出来、$m$-次パラ $W_N$ 代数という恐ろしい代数が出てくると思われています\cite{Nishioka:2011jk}。特に $m=N=2$ の場合は、$2$-次 パラ$W_2$ 代数というのは通常の超対称ビラソロ代数になるので具体的な確認をいくつもすることが出来、最近いろいろと論文が出ています\cite{Belavin:2011pp,Bonelli:2011jx,Bonelli:2011kv}。

\section*{謝辞}
まずは夏の学校「数理物理2011」の世話人の皆様に、このような機会を与えてくださった事を感謝したいと思います。
著者の仕事は部分的にアメリカNSFのグラント番号 PHY-0969448 及び高等研究所の Marvin L. Goldberger membership からの援助を受けました。また、数物連携宇宙機構を通じて日本国文部科学省世界トップレベル研究拠点プログラムからの補助も受けています。

\bibliographystyle{ytphys}
\def\refname{参考文献}
\bibliography{ref}

\providecommand{\href}[2]{#2}\begingroup\raggedright\begin{thebibliography}{10}

\bibitem{Gakkaishi}
{立川裕二}, ``共形場理論とインスタントンの統計力学,''
  {\em 物理学会誌} {\bfseries 65(9)} (2010) 703--707.

\bibitem{ID}
C.~Itzykson and J.-M. Druffe, {\em Statistical Field Theory}.
\newblock Cambridge University Press, 1991.

\bibitem{KY}
{川上則雄・梁成吉}, {\em 共形場理論と一次元量子系}.
\newblock 岩波書店, 1997.

\bibitem{Yamada}
{山田泰彦}, {\em 共形場理論入門}.
\newblock 培風館, 2006.

\bibitem{ItoText}
{伊藤克司}, {\em 共形場理論}.
\newblock サイエンス社, 2011.

\bibitem{Marshakov:2009gn}
A.~Marshakov, A.~Mironov, and A.~Morozov, ``{On Non-Conformal Limit of the AGT
  Relations},'' \href{http://dx.doi.org/10.1016/j.physletb.2009.10.077}{{\em
  Phys. Lett.} {\bfseries B682} (2009) 125--129},
\href{http://arxiv.org/abs/0909.2052}{{\ttfamily arXiv:0909.2052 [hep-th]}}.

\bibitem{Coleman}
S.~Coleman, {\em Aspects of Symmetry}.
\newblock Cambridge University Press, 1988.

\bibitem{tHooft:1999au}
G.~'t~Hooft, ``{Monopoles, Instantons and Confinement},''
\href{http://arxiv.org/abs/hep-th/0010225}{{\ttfamily arXiv:hep-th/0010225}}.

\bibitem{Atiyah:1978ri}
M.~F. Atiyah, N.~J. Hitchin, V.~G. Drinfeld, and Y.~I. Manin, ``{Construction
  of Instantons},''
\href{http://dx.doi.org/10.1016/0375-9601(78)90141-X}{{\em Phys. Lett.}
  {\bfseries A65} (1978) 185--187}.

\bibitem{Dorey:2002ik}
N.~Dorey, T.~J. Hollowood, V.~V. Khoze, and M.~P. Mattis, ``{The Calculus of
  Many Instantons},''
  \href{http://dx.doi.org/10.1016/S0370-1573(02)00301-0}{{\em Phys. Rept.}
  {\bfseries 371} (2002) 231--459},
\href{http://arxiv.org/abs/hep-th/0206063}{{\ttfamily arXiv:hep-th/0206063}}.

\bibitem{Ito}
{伊藤克司}, ``{$\cN=2$ 超対称ゲージ理論とインスタントン},''
  \href{{http://www2.yukawa.kyoto-u.ac.jp/~yonupa/particle/2004/Lecture_Note.pdf}}{{\em
  素粒子論研究} {\bfseries 116(4)} (2008) 5--107}.

\bibitem{'tHooft:1976fv}
G.~'t~Hooft, ``{Computation of the Quantum Effects Due to a Four- Dimensional
  Pseudoparticle},''
\href{http://dx.doi.org/10.1103/PhysRevD.14.3432}{{\em Phys. Rev.} {\bfseries
  D14} (1976) 3432--3450}.

\bibitem{GS}
V.~Guillemin and S.~Sternberg, {\em Symplectic Techniques in Physics}.
\newblock Cambridge University Press, 1990.

\bibitem{NakajimaSougou}
{中島啓},
  ``インスタントンの数え上げとドナルドソン不変量,''
  \href{{http://www.kurims.kyoto-u.ac.jp/~nakajima/Talks/sougou2011.pdf}}{{\em
  日本数学会年会アブストラクト集} (2011) }.

\bibitem{Moore:1997dj}
G.~W. Moore, N.~Nekrasov, and S.~Shatashvili, ``{Integrating over Higgs
  Branches},'' \href{http://dx.doi.org/10.1007/PL00005525}{{\em Commun. Math.
  Phys.} {\bfseries 209} (2000) 97--121},
\href{http://arxiv.org/abs/hep-th/9712241}{{\ttfamily arXiv:hep-th/9712241}}.

\bibitem{Nekrasov:2002qd}
N.~A. Nekrasov, ``{Seiberg-Witten Prepotential from Instanton Counting},'' {\em
  Adv. Theor. Math. Phys.} {\bfseries 7} (2004) 831--864,
\href{http://arxiv.org/abs/hep-th/0206161}{{\ttfamily arXiv:hep-th/0206161}}.

\bibitem{Flume:2002az}
R.~Flume and R.~Poghossian, ``{An Algorithm for the Microscopic Evaluation of
  the Coefficients of the Seiberg-Witten Prepotential},''
  \href{http://dx.doi.org/10.1142/S0217751X03013685}{{\em Int. J. Mod. Phys.}
  {\bfseries A18} (2003) 2541},
\href{http://arxiv.org/abs/hep-th/0208176}{{\ttfamily arXiv:hep-th/0208176}}.

\bibitem{Nakajima:2003uh}
H.~Nakajima and K.~Yoshioka, ``{Lectures on Instanton Counting},''
\href{http://arxiv.org/abs/math/0311058}{{\ttfamily arXiv:math/0311058}}.

\bibitem{Alday:2009aq}
L.~F. Alday, D.~Gaiotto, and Y.~Tachikawa, ``{Liouville Correlation Functions
  from Four-Dimensional Gauge Theories},''
  \href{http://dx.doi.org/10.1007/s11005-010-0369-5}{{\em Lett. Math. Phys.}
  {\bfseries 91} (2010) 167--197},
\href{http://arxiv.org/abs/0906.3219}{{\ttfamily arXiv:0906.3219 [hep-th]}}.

\bibitem{Gaiotto:2009ma}
D.~Gaiotto, ``{Asymptotically Free ${\mathcal{N}}\!=2$ Theories and Irregular
  Conformal Blocks},''
\href{http://arxiv.org/abs/0908.0307}{{\ttfamily arXiv:0908.0307 [hep-th]}}.

\bibitem{Poghossian:2009mk}
R.~Poghossian, ``{Recursion Relations in CFT and ${\mathcal{N}}\!=2$ SYM
  Theory},'' \href{http://dx.doi.org/10.1088/1126-6708/2009/12/038}{{\em JHEP}
  {\bfseries 12} (2009) 038},
\href{http://arxiv.org/abs/0909.3412}{{\ttfamily arXiv:0909.3412 [hep-th]}}.

\bibitem{Fateev:2009aw}
V.~A. Fateev and A.~V. Litvinov, ``{On AGT Conjecture},''
  \href{http://dx.doi.org/10.1007/JHEP02(2010)014}{{\em JHEP} {\bfseries 02}
  (2010) 014},
\href{http://arxiv.org/abs/0912.0504}{{\ttfamily arXiv:0912.0504 [hep-th]}}.

\bibitem{Hadasz:2010xp}
L.~Hadasz, Z.~Jask\'olski, and P.~Suchanek, ``{Proving the AGT Relation for
  $N_F = 0,1,2$ Antifundamentals},''
\href{http://arxiv.org/abs/1004.1841}{{\ttfamily arXiv:1004.1841 [hep-th]}}.

\bibitem{Nakajima}
H.~Nakajima, ``{Instantons on ALE Spaces, Quiver Varieties and Kac-Moody
  Algebras},'' \href{http://dx.doi.org/10.1215/S0012-7094-94-07613-8}{{\em Duke
  Math. Journal} {\bfseries 76} (1994) 365}.

\bibitem{MO}
D.~Maulik and A.~Okounkov, {\em unpublished}.
\newblock 2011?

\bibitem{Witten:2009at}
E.~Witten, ``{Geometric Langlands from Six Dimensions},''
\href{http://arxiv.org/abs/0905.2720}{{\ttfamily arXiv:0905.2720 [hep-th]}}.

\bibitem{Witten:2009mh}
E.~Witten, ``{Geometric Langlands and the Equations of Nahm and Bogomolny},''
\href{http://arxiv.org/abs/0905.4795}{{\ttfamily arXiv:0905.4795 [hep-th]}}.

\bibitem{Dijkgraaf:1997vv}
R.~Dijkgraaf, E.~P. Verlinde, and H.~L. Verlinde, ``{Matrix String Theory},''
  \href{http://dx.doi.org/10.1016/S0550-3213(97)00326-X}{{\em Nucl. Phys.}
  {\bfseries B500} (1997) 43--61},
\href{http://arxiv.org/abs/hep-th/9703030}{{\ttfamily arXiv:hep-th/9703030}}.

\bibitem{Berkooz:1997cq}
M.~Berkooz, M.~Rozali, and N.~Seiberg, ``{Matrix Description of M Theory on $
  T^4$ and $ T^5$},''
  \href{http://dx.doi.org/10.1016/S0370-2693(97)00800-9}{{\em Phys. Lett.}
  {\bfseries B408} (1997) 105--110},
\href{http://arxiv.org/abs/hep-th/9704089}{{\ttfamily arXiv:hep-th/9704089}}.

\bibitem{Polchinski}
J.~Polchinski, {\em String Theory}.
\newblock Cambridge University Press, 1998.

\bibitem{Bouwknegt:1992wg}
P.~Bouwknegt and K.~Schoutens, ``{W Symmetry in Conformal Field Theory},''
  \href{http://dx.doi.org/10.1016/0370-1573(93)90111-P}{{\em Phys. Rept.}
  {\bfseries 223} (1993) 183--276},
\href{http://arxiv.org/abs/hep-th/9210010}{{\ttfamily arXiv:hep-th/9210010}}.

\bibitem{Wakimoto}
脇本実, ``{Ｗ代数の話 $\sim$
  スーパー・コンフォーマル代数の表現論へ},'' in {\em
  {「数理物理2004」予稿集}}.
\newblock 2004.

\bibitem{Wyllard:2009hg}
N.~Wyllard, ``{$A_{N-1}$ Conformal Toda Field Theory Correlation Functions from
  Conformal ${\mathcal{N}}\!=2$ $SU(N)$ Quiver Gauge Theories},''
  \href{http://dx.doi.org/10.1088/1126-6708/2009/11/002}{{\em JHEP} {\bfseries
  11} (2009) 002},
\href{http://arxiv.org/abs/0907.2189}{{\ttfamily arXiv:0907.2189 [hep-th]}}.

\bibitem{Mironov:2009by}
A.~Mironov and A.~Morozov, ``{On AGT Relation in the Case of U(3)},''
\href{http://arxiv.org/abs/0908.2569}{{\ttfamily arXiv:0908.2569 [hep-th]}}.

\bibitem{Taki:2009zd}
M.~Taki, ``{On AGT Conjecture for Pure Super Yang-Mills and W- algebra},''
  \href{http://dx.doi.org/10.1007/JHEP05(2011)038}{{\em JHEP} {\bfseries 05}
  (2011) 038},
\href{http://arxiv.org/abs/0912.4789}{{\ttfamily arXiv:0912.4789 [hep-th]}}.

\bibitem{Braverman:2010ef}
A.~Braverman, B.~Feigin, M.~Finkelberg, and L.~Rybnikov, ``{A Finite Analog of
  the AGT Relation I: Finite W-Algebras and Quasimaps' Spaces},''
\href{http://arxiv.org/abs/1008.3655}{{\ttfamily arXiv:1008.3655 [math.AG]}}.

\bibitem{Wyllard:2010vi}
N.~Wyllard, ``{Instanton Partition Functions in ${\mathcal{N}}\!=2$ $SU(N)$
  Gauge Theories with a General Surface Operator, and Their W-Algebra Duals},''
  \href{http://dx.doi.org/10.1007/JHEP02(2011)114}{{\em JHEP} {\bfseries 02}
  (2011) 114},
\href{http://arxiv.org/abs/1012.1355}{{\ttfamily arXiv:1012.1355 [hep-th]}}.

\bibitem{Kanno:2011fw}
H.~Kanno and Y.~Tachikawa, ``{Instanton Counting with a Surface Operator and
  the Chain- Saw Quiver},''
  \href{http://dx.doi.org/10.1007/JHEP06(2011)119}{{\em JHEP} {\bfseries 06}
  (2011) 119},
\href{http://arxiv.org/abs/1105.0357}{{\ttfamily arXiv:1105.0357 [hep-th]}}.

\bibitem{Nishioka:2011jk}
T.~Nishioka and Y.~Tachikawa, ``{Para-Liouville/Toda Central Charges from
  M5-Branes},''
\href{http://arxiv.org/abs/1106.1172}{{\ttfamily arXiv:1106.1172 [hep-th]}}.

\bibitem{Belavin:2011pp}
V.~Belavin and B.~Feigin, ``{Super Liouville Conformal Blocks from
  ${\mathcal{N}}\!=2$ $SU(2)$ Quiver Gauge Theories},''
  \href{http://dx.doi.org/10.1007/JHEP07(2011)079}{{\em JHEP} {\bfseries 07}
  (2011) 079},
\href{http://arxiv.org/abs/1105.5800}{{\ttfamily arXiv:1105.5800 [hep-th]}}.

\bibitem{Bonelli:2011jx}
G.~Bonelli, K.~Maruyoshi, and A.~Tanzini, ``{Instantons on ALE Spaces and Super
  Liouville Conformal Field Theories},''
\href{http://arxiv.org/abs/1106.2505}{{\ttfamily arXiv:1106.2505 [hep-th]}}.

\bibitem{Bonelli:2011kv}
G.~Bonelli, K.~Maruyoshi, and A.~Tanzini, ``{Gauge Theories on ALE Space and
  Super Liouville Correlation Functions},''
\href{http://arxiv.org/abs/1107.4609}{{\ttfamily arXiv:1107.4609 [hep-th]}}.

\end{thebibliography}\endgroup

\ifxetex\else\end{CJK}\fi
\end{document}